\documentclass[iop]{emulateapj}
\usepackage{natbib}

\usepackage{ulem,footnote,lineno,graphicx,color,rotating,amssymb,amsmath,amsfonts,times,hyperref, booktabs,soul}

\newcommand{\muG}{\mu{\rm G}}

\newcommand{\lsim}{\mathrel{\mathop{\kern 0pt \rlap
  {\raise.2ex\hbox{$<$}}}
  \lower.9ex\hbox{\kern-.190em $\sim$}}}
\newcommand{\gsim}{\mathrel{\mathop{\kern 0pt \rlap
  {\raise.2ex\hbox{$>$}}}
  \lower.9ex\hbox{\kern-.190em $\sim$}}}

\begin{document}

\preprint[TUM-HEP-1078/17

\title{Theoretical interpretation of Pass 8  {\it Fermi}-LAT \lowercase{$e^+ + e^-$} data}
\author{M. Di Mauro$^1$, S. Manconi$^{2,3}$,  A. Vittino$^4$,  F. Donato$^{2,3}$, N. Fornengo$^{2,3}$,
L.~Baldini\altaffilmark{5}, 
R.~Bonino\altaffilmark{2,3}, 
N.~Di~Lalla\altaffilmark{5}, 
L.~Latronico\altaffilmark{3}, 
S.~Maldera\altaffilmark{3}, 
A.~Manfreda\altaffilmark{5}, 
M.~Negro\altaffilmark{2,3}, 
M.~Pesce-Rollins\altaffilmark{6}, 
C.~Sgr\`o\altaffilmark{6}, 
F.~Spada\altaffilmark{6}}

\affil{$^1$W. W. Hansen Experimental Physics Laboratory, Kavli Institute for Particle Astrophysics and Cosmology, Department of Physics and SLAC National Accelerator Laboratory, Stanford University, Stanford, CA 94305, USA}
\affil{$^2$Department of Physics, University of Torino, via P. Giuria 1, 10125 Torino, Italy}
\affil{$^3$Istituto Nazionale di Fisica Nucleare, via P. Giuria 1, 10125 Torino, Italy}
\affil{$^4$Physik-Department T30D, Technische Universit\"{a}t M\"{u}nchen, James-Franck Stra{\ss}e 1, D-85748 Garching, Germany}
\affil{$^5$ Universit\`a di Pisa and Istituto Nazionale di Fisica Nucleare, Sezione di Pisa I-56127 Pisa, Italy}
\affil{$^6$Istituto Nazionale di Fisica Nucleare, Sezione di Pisa, I-56127 Pisa, Italy}

\def \aap  {A\&A}
\def \aaps  {A\&AS}
\def \aj  {AJ}
\def \apj  {ApJ}
\def \apjs  {ApJS}
\def \apjl  {ApJL}
\def \apss  {AP\&SS}
\def \araa  {ARA\&A}
\def \jcap  {JCAP}
\def \prd {Phy.Rev.D}
\def \ssr {SSRv}
\def \mnras {MNRAS}
\def \nat {Nature}
\def \physrep {Phys.Rept.}
\def \pasj {PASJ}
\def \etal {et~al.~}
\def \rmxaa {RMxAA}
\def \jgr {JGR}

\begin{abstract} 
The flux of positrons and electrons ($e^+ + e^-$) has been measured by the {\it Fermi} Large Area Telescope (LAT) in the energy range between 7~GeV and 2~TeV. 
We discuss a number of interpretations of Pass 8 {\it Fermi}-LAT \lowercase{$e^+ + e^-$} spectrum, combining electron and positron emission from supernova remnants (SNRs) and pulsar wind nebulae (PWNe), or produced by the collision of cosmic rays with the interstellar medium. 
We find that the {\it Fermi}-LAT spectrum is compatible with the sum of electrons from a smooth SNR population,  positrons from cataloged PWNe, and a secondary component. 
{ If} we include { in our analysis} constraints from AMS-02 positron spectrum, { we} obtain a slightly worse fit to the $e^+ + e^-$ {\it Fermi}-LAT spectrum, depending on the propagation model.
As an additional scenario, we replace the smooth SNR component within 0.7 kpc with the { individual sources} found in Green's catalog of Galactic SNRs. { We find that separate consideration of far and near sources helps to reproduce} the $e^+ + e^-$ {\it Fermi}-LAT spectrum. { However, we show that the fit degrades when the radio constraints on the positron emission from Vela SNR (which is the main contributor at high energies) are taken into account. }
We find that a break in the power-law injection spectrum at about 100 GeV {can also reproduce the measured $e^+ + e^-$ spectrum}  and, among the cosmic-ray propagation models that we consider, no reasonable break of the power-law dependence of the diffusion coefficient can modify the electron flux enough to reproduce the observed shape. 
\end{abstract}


\maketitle

\section{Introduction}

Investigation of the leptonic component of cosmic rays (CRs) provides invaluable insight into the properties of CR sources and CR propagation. At present, the most accurate measurements of the different observables related to CR leptons have been performed by the AMS-02 experiment \citep{2014PhRvL.113l1101A, 2014PhRvL.113l1102A,Aguilar:2014fea}. The data provided by AMS-02 have been interpreted within several theoretical models: e.g., \cite{Blum:2013zsa} discuss the possibility of a purely secondary origin of positrons, while \cite{Bergstrom:2013jra,Gaggero:2013nfa,Mertsch:2014poa, Delahaye:2014osa,Jin:2014ica,Ibarra:2013zia, Boudaud:2014dta,  Lin:2014vja,   Yuan:2013eja} and \cite{DiMauro:2015jxa} investigate the properties of additional positron sources (pulsars, dark matter or acceleration within supernovae). Furthermore, CR leptons have been investigated in connection with other observables, such as hadronic CR fluxes \citep{ Tomassetti:2015cva, Tomassetti:2015mha, Kachelriess:2015oua,2016arXiv160802018L} or synchrotron emission across the Galaxy \citep{2013JCAP...03..036D, 2013MNRAS.436.2127O, 2016arXiv160100546P}.


The {\it Fermi}-LAT Collaboration has recently reported a new measurement of the inclusive CR positron and electron ($e^+ + e^-$) spectrum between 7~GeV and 2~TeV, obtained with almost seven years of Pass 8 data \citep{Abdollahi:2017nat,PhysRevLett.118.091103}. The LAT spectrum suggests the presence of a break at about 50 GeV but this feature is not statistically significant when the systematic uncertainty on the energy measurement is taken into account. In this work we choose to use the new LAT spectrum without taking into account this specific uncertainty. In that case, a fit to the LAT spectrum between 7 GeV and 2 TeV with a broken power-law is reported to yield a break at $E_b=(53\pm8)$~GeV with spectral indices below and above the break $\gamma=(3.21\pm0.02)$ and $\gamma=(3.07\pm0.02)$.
We refer to \cite{PhysRevLett.118.091103} for a discussion of the difference between AMS-02 and {\it Fermi}-LAT $e^+ + e^-$ spectrum that is  at the level of $1.7\sigma$ for $E>30$ GeV.

Here we study the {\it Fermi}-LAT results, including the potential new feature of a spectral break, within the theoretical model proposed in \cite{DiMauro:2014iia,DiMauro:2015jxa} and \cite{Manconi:2016byt}, which has already been used to study the AMS-02 electron and positron spectra. 
In this model, electrons and positrons are either emitted by primary astrophysical sources, such as supernova remnants (SNRs) and pulsar wind nebulae (PWNe), or they are produced as a secondary CR component, due to collisions of protons and helium nuclei with the interstellar medium (ISM). 

The paper is organized as follows: Section~\ref{sec:contributions} describes the different contributions to the $e^+ + e^-$ flux, while Section~\ref{sec:transport} illustrates the model that we use for the propagation of electrons and positrons through the Galaxy. Sections~\ref{sec:results} and \ref{sec:breaks} discuss our analysis and results and we conclude in Section~\ref{sec:conclusions}.

\medskip
\section{Contributions to the \lowercase{$e^+ + e^-$} flux}
\label{sec:contributions}

Electrons and positrons can be products of a variety of processes that take place in the Galaxy. In this Section we briefly outline of the different production mechanisms and we describe our modeling.  More details can be found in  ~\cite{2010A&A...524A..51D,DiMauro:2014iia,DiMauro:2015jxa} and \cite{Manconi:2016byt}.

\subsection{Supernova Remnants}
\label{subsec:SNRs}

SNRs are commonly considered as main accelerators of Galactic CRs. Charged particles scatter repeatedly upstream and downstream of the shock wave that is generated by the stellar explosion and receive an increase in energy each time they cross the shock front. This mechanism of diffusive shock acceleration produces a spectrum of accelerated particles that we assume to be described in terms of a power law with an exponential cut-off:
 
\begin{equation}
     \label{eq:Q_SNR}
         Q(E)=Q_{0, \rm SNR} \left(\frac{E}{E_0}\right)^{-\gamma_{\rm SNR}} e^{-\frac{E}{E_c}}.
\end{equation}
where $E_0$=1 GeV. The injection spectrum in Equation \eqref{eq:Q_SNR} is related to the total energy (in units of erg or GeV) emitted in electrons by SNRs (analogously can be written for electrons and positrons by PWN)
\begin{equation}
E_{tot,{\rm SNR}} =  \int^{\infty}_{E_{\rm{min}}} E Q(E) dE, 
\label{eq:eleenergy}
\end{equation}
where we fix $E_{\rm{min}}$=0.1 GeV.
We fix the average Galactic supernova explosion rate (usually indicated with $\Gamma_*$) to 1/century.
The spectrum of particles accelerated by SNRs is therefore completely described by three parameters: the normalization 
$Q_{0, \rm SNR}$ or, equivalently $E_{tot,{\rm SNR}}$, the spectral index $\gamma_{\rm SNR}$ and the cut-off energy $E_c$. As mentioned in \cite{DiMauro:2014iia}, radio measurements in the SNR region can provide insight into the values of $E_{tot,{\rm SNR}}$ through the 
magnetic field, and $\gamma_{\rm SNR}$. As for the cut-off energy $E_c$, both theoretical considerations and observational evidence place it in the multi-TeV range.  
Radio and gamma-ray observations indicate that the energy cutoff should be in the TeV range \citep[see, e.g.,][]{0004-637X-525-1-368,Aharonian:2001mz,Aharonian:2008aa,Aharonian:2008nw,2010ApJ...714..163A}.
Throughout this paper, we assume $E_c$= 5 TeV, for the acceleration both by SNRs as well as PWNe (Equation \eqref{eq:Q_PWN}). 
One important aspect is that SNRs accelerate particles that are already present in the environment of the explosion, namely the ISM: since in the ISM electrons are much more abundant than positrons, SNRs can be considered to accelerate only electrons. 

As explained in detail in \cite{DiMauro:2014iia}, for the purposes of our analysis we divide the SNRs into two categories.
We define here $R \equiv |\bold{r} - \bold{r}_{\odot}|$, where $r$ is the Galactocentric radial coordinate along the Galactic plane and $r_{\odot}$ is the solar position. 
According to their distance from the Earth, we consider:

\begin{itemize}
\item{Far SNRs ($R >  R_{\rm cut}$): they are treated as a population of sources that are spatially distributed according to the \cite{2004IAUS..218..105L} (hereafter L04) or \cite{2015MNRAS.454.1517G} (G15) distributions.
The spatial distribution G15 is a new estimation based on most-recent Galactic SNR catalog \citep{Green:2014cea}. 
The distribution L04 is for pulsars that can be used as tracers of the SNR distribution.
Far SNRs are assumed to contribute to CR electron production as in Equation (\ref{eq:Q_SNR}), with a common normalization $E_{tot,{\rm SNR}}$ and spectral index $\gamma_{\rm SNR}$. 
These are usually taken to be free parameters in our fits.}
\item{Near SNRs ($R \leq  R_{\rm cut}$): these sources are taken from Green's catalog \citep{Green:2014cea}, which provides information on their distance, age, magnetic field B and $\gamma_{\mathrm{SNR}}$. As in  \cite{DiMauro:2015jxa} and \cite{Manconi:2016byt}, we allow separate free normalization of the flux generated by the Vela SNR, which is the most powerful source among the nearby SNRs.}
\end{itemize} 
We can treat the cut as a cylinder centered at Earth, with radius equal to $ R_{\rm cut}$. More details are given in Section \ref{sec:distributions}. 
The motivation to separate SNRs into near and far components is due to the fact that far SNRs contribute to the electron flux mostly at low energies, while local sources likely dominate the high-energy tail. For the latter, we have more specific information from Green's catalog, which lets us treat the nearby SNR component as individual sources with physical parameters based on observations at some wavelength (mostly radio). This allows us to investigate in greater depth the high-energy portion of the $e^+ + e^-$ spectrum. We perform dedicated analyses for different choices of the
$ R_{\rm cut}$ parameter, including $ R_{\rm cut} = 0$, which extends the average distribution of the SNR population to the whole Galaxy. In this case clearly no catalog sources are included. 

\subsection{Pulsar Wind Nebulae}
\label{subsec:PWNe} 

Pulsars can produce a flux of electrons and positrons. As described, e.g.,~in \cite{1970ApJ...162L.181S,Ruderman:1975ju,1976ApJ...203..209C,Cheng:1986qt,1987ICRC....2...92H,1996SSRv...75..235A,2001AA...368.1063Z} and \cite{Amato:2013fua},  electrons stripped from the neutron star surface generate pairs cascades in the pulsar magnetosphere. These electrons and positrons are further accelerated in the pulsar's stripped wind and/or at its termination shock, beyond which a nebula of very high energy particles forms: the PWN.
As for SNRs, we assume that the energy spectrum of electrons and positrons emitted by a PWN can be described as a power-law with an exponential cut-off:

\begin{equation}
     \label{eq:Q_PWN}
         Q(E)=Q_{0, \rm PWN} \left(\frac{E}{E_0}\right)^{-\gamma_{\rm PWN}} e^{-\frac{E}{E_c}}.
\end{equation}  
In our modeling, we express the normalization of the PWN spectrum $Q_{0, \rm PWN}$ in terms of the spin-down energy of the pulsar $W_0$, which is the energy emitted by the pulsar as it slows down \citep{Hooper:2008kg}:
\begin{equation}
W_0\approx\tau_0\dot{E}\left(1+\frac{t_*}{\tau_0}\right)^2.
\end{equation}
where $t_*$ is the present age of the pulsar, $\tau_0$ is the typical pulsar decay time and $\dot{E}$ is the spin-down luminosity. The normalization $Q_{0, \rm PWN}$ is therefore obtained from the relation:

\begin{equation}
\label{Wop}
\int_{E_{\rm min}}^{\infty}\,dE\,E\,Q(E)\,=\eta_{\rm PWN}\,W_0,
\end{equation}
where $\eta_{\rm PWN}$ is the efficiency factor for the conversion of the spin-down energy into electrons and positrons.

As in~\cite{DiMauro:2014iia,DiMauro:2015jxa} and \cite{Manconi:2016byt}, we consider in our analysis all the pulsars in the continuously updated ATNF catalog\footnote{\url{http://www.atnf.csiro.au/people/pulsar/psrcat/}. We use catalog version 1.55}, which provides the spin-down energy, age and distance of each PWN.
For definiteness, we select only PWNe with ages greater than $50$~kyr. This is based on the fact that the release of electron and positron pairs in the ISM is estimated to occur at least $40-50$~kyr after the formation of the pulsar \citep{2011ASSP...21..624B}.  
The efficiency $\eta_{\rm PWN}$ and the spectral index $\gamma_{\rm PWN}$  are our free parameters. 

\subsection{Secondary positrons and electrons}
\label{subsec:secondaries}

Electrons and positrons of secondary origin are produced by interactions of primary CR nuclei with the gas nuclei in the ISM. The source term for this contribution is: 
\begin{eqnarray}
\label{eq:source}
  Q_{\mathrm{sec}}(\mathbf{x},E_{e}&) &= 4 \pi \; \sum_{i,j}  \displaystyle \int  \Phi_{{\rm CR},i} \left( \mathbf{x} , E_{\rm CR} \right) \nonumber
  \\
&& \left. \frac{d\sigma}{dE_{e}}\right|_{i,j}(E_{\rm CR}, E_{e}) n_{{\rm ISM},j} \,\,  dE_{\rm CR} ,
\end{eqnarray}
where $i$ runs over the primary CR species of flux density $\Phi_{{\rm CR},i}$ and $j$ over the target nuclei in the ISM of density $n_{{\rm ISM},j}$  considered constant with $n_H = 0.9$ cm$^{-3}$ for Hydrogen and $n_{He} = 0.1$ cm$^{-3}$ for Helium. $n_{\rm ISM}$ is confined in a thin disk of half-height 100 pc \cite[see, e.g.,][and references therein]{Delahaye:2008ua}.
$\mathbf{x}$ is the position vector in the Galaxy and  $d\sigma / dE_{e}$ is the differential cross section for electron and positron production in the spallation reaction under consideration \citep{2006ApJ...647..692K}. 

We determine the source term of Equation (\ref{eq:source}) following the same approach detailed in \cite{DiMauro:2014iia} and \cite{DiMauro:2015jxa}, where we adopt primary CR fluxes obtained by fitting the AMS-02 data on protons and helium. The parameters of the spectra determined in this fit are reported in  \cite{DiMauro:2015jxa}.

\section{Transport of charged particles in the Galaxy}
\label{sec:transport}

Electron and positron transport in the Galaxy is treated by means of a semi-analytical model, following the same approach as \cite{2010A&A...524A..51D} and \cite{DiMauro:2014iia}. 
The semi-analytical model is simplified compared to codes like GALPROP \citep{Moskalenko:1997gh,Strong:1998pw,2011CoPhC.182.1156V} and DRAGON \citep{Evoli:2008dv,Evoli:2016xgn}; by numerically solving the transport equation, these codes can implement more complex features of the Galactic environment and its geometry (e.g. Galaxy spiral arms or small scale inhomogeneities). 

However, we expect such complex features to have at most a mild impact on the problem at hand. blueThis is for example the case of implementing spiral arms in the distribution of SNRs, as we discuss in Appendix \ref{ap:sp} and in Section \ref{sec:smooth}. In fact, as discussed in \cite{2010A&A...524A..51D}, due to energy losses, both primary and secondary leptons that reach Earth are produced predominantly within a few kpc from the Sun. In this small region of the Galaxy, the considerations for evaluating CR transport (e.g., the magnetic field, the interstellar radiation field and the diffusion coefficient) are unlikely to have a strong spatial dependence and therefore our semi-analytical approach is an acceptable approximation. In addition, relative to fully numerical methods, the semi-analytical model has faster execution times and allows for larger parameter-space scans. 

Let us provide a brief summary of the model we employ. For details, we refer to \cite{2010A&A...524A..51D}. Independent of the production mechanism, charged CRs propagate through the Galactic magnetic field irregularities and experience a number of different physical processes. 
CRs are confined by Galactic magnetic fields of mean value $B\sim 1-5\mu$G in a propagation zone called the diffusive halo, which we model as a thick disk which matches the structure of our Galaxy. 
The radial extension of the disc is fixed to $r_{\rm disc}=20$~kpc, while its vertical half height is quite uncertain, $L\simeq 1$-$15$~kpc.  
The electron number density per unit energy  $\psi= \psi(E, \mathbf{x}, t)$ is linked to the electron flux $\Phi = \frac{v}{4\pi} \psi$, where $v$ is the electron velocity (de facto $v$=c). 
The transport of electrons with energy $E$ in the diffusive halo is described through the transport equation:
\begin{equation}
\label{eq:transport}
 \partial_t \psi  - \mathbf{\nabla} \cdot \left\lbrace K(E)  \mathbf{\nabla} \psi \right\rbrace + \partial_E \left\lbrace b(E) \psi \right\rbrace = Q(E, \mathbf{x}, t)
\end{equation}
which accounts for the main processes that charged leptons experience while propagating to the Earth. 
Above a few GeV of energy the propagation of electrons is dominated by spatial diffusion, parameterized through a diffusion coefficient $K(E)$, and energy losses $b(E)$. Specifically, synchrotron emission and inverse Compton (IC) scattering dominate over ionization, adiabatic and bremsstrahlung energy losses \citep[see, e.g.,][]{Delahaye:2008ua}. Diffusion in momentum space due to motions of the turbulent magnetic field, as well as the effect of the Galactic convective wind, are sub-dominant for electrons that reach the Earth with energies $E \gtrsim 5$~GeV \citep{Delahaye:2008ua}. We recall that in our model a fully relativistic description of IC energy losses and a mean value of $B_{sync}= 3.6\,\mu$G are used  \citep{Delahaye:2008ua}. The diffusion coefficient can be in general a function of position in the Galaxy $K(E, \mathbf{x})$, as done e.g. in 
\cite{Tomassetti:2015mha}. However, the propagation scale (see Equation \eqref{eq:lambda} below) for high-energy electrons is a few kpc \citep{2010A&A...524A..51D}. Furthermore, the diffusion structure of our Galaxy is still not well known.  We therefore assume a spatially uniform $K(E)$ throughout the diffusive halo, which permits a full semi-analytical solution:
\begin{equation}
 K(E)= \beta K_0 (R/1 \text{GV})^\delta \simeq K_0 (E/1 \text{GeV})^{\delta}
 \label{eq:K}
\end{equation}
where the right-hand side is valid because the rigidity of electrons is $R\sim E$ and $\beta \simeq 1$ at the energies under consideration.

The propagation parameters ($\delta$, $K_0$, $L$) are generally constrained by means of the secondary-to-primary ratio B/C computed within the same model and confronted with CR data. Specifically, we will use the MED and MAX sets of parameters \citep{Donato:2003xg} ($\delta=0.70$, $K_0=0.0112$ kpc$^2$/Myr and $L=4$ kpc for MED and $\delta=0.46$, $K_0=0.0765$ kpc$^2$/Myr and $L=15$ kpc for MAX), since the MIN model has been disfavored by studies of positrons at low energies \citep{Lavalle:2014kca}. We also verified that the new parameter sets recently obtained by \cite{Kappl:2015bqa} (hereafter Kappl2015) and \cite{Genolini:2015cta} (hereafter Genolini2015) from the preliminary AMS-02 B/C data \citep{amsbcration} give electron fluxes that fall between our MED and MAX results. 
As we will discuss in the sections of the results, using other propagation models would slightly modify the values of some parameters in our models without however changing our conclusions.

Equation \eqref{eq:transport} is solved according to the semi-analytical model extensively described in \cite{2010A&A...524A..51D} and \cite{DiMauro:2014iia}. The solutions for a smooth and steady distribution of sources and for a discrete and time-dependent case are outlined in the next two subsections. The solutions for secondary electrons and positrons is computed as described in  \cite{Delahaye:2008ua}, to which we refer for details.

\subsection{Smooth distribution of sources in the Galaxy}
\label{sec:distributions}

\begin{figure}[t]
\centering
\includegraphics[width=\columnwidth]{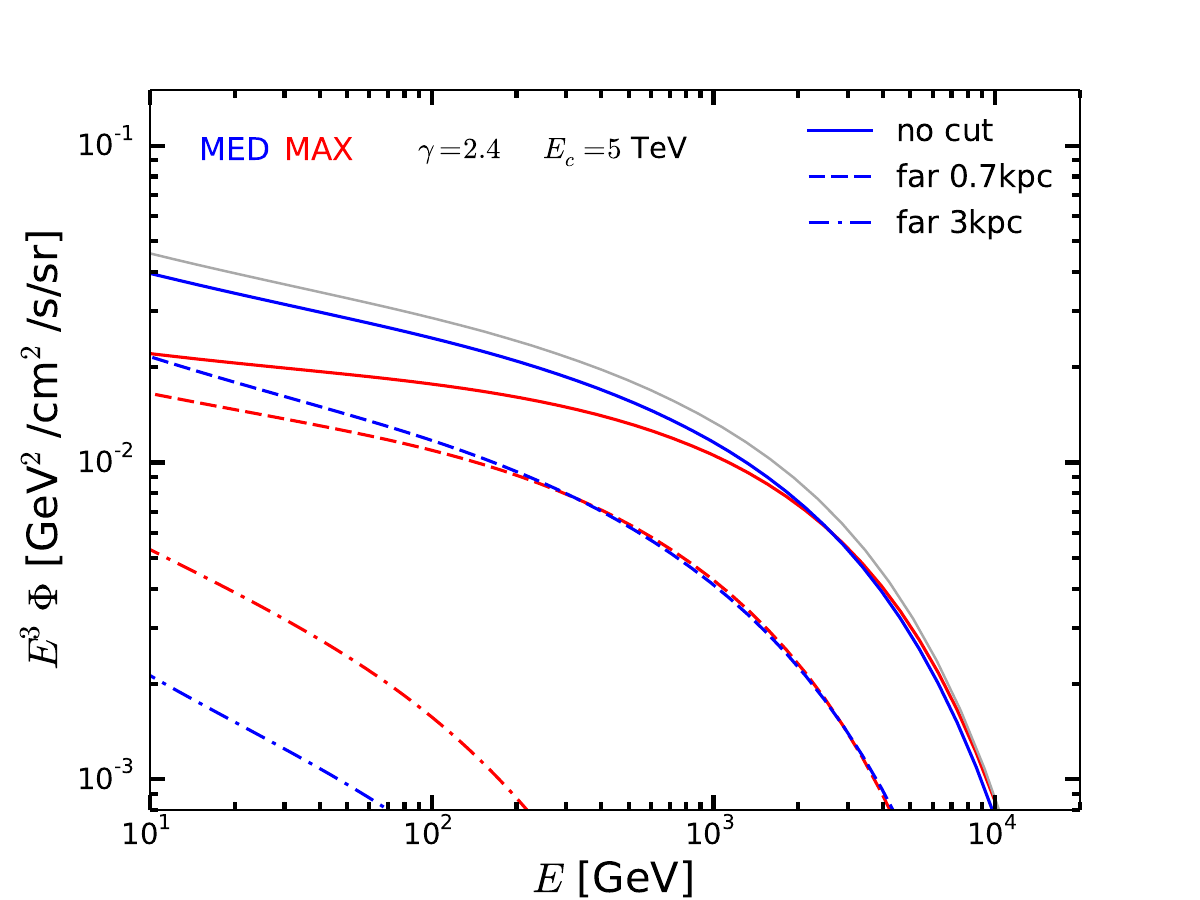}
\caption{Electron fluxes at Earth, originated from a smooth SNR distribution with different cuts on distance from the Sun $ R_{\rm cut}$, propagation models and SNR spatial distribution models. Blue (red) lines refer to the MED (MAX) propagation model, while different styles show the results when a cut on the distribution of SNRs is applied: dot-dashed shows the results when only far SNRs are considered ($R> R_{\rm cut} = 3$ kpc),  dashed stands for $R> R_{\rm cut} = 0.7$ kpc and solid shows the case when the full smooth SNR distribution is taken ($ R_{\rm cut} = 0$). In all cases, the SNR distribution is G15.
The solid pale grey line shows the result for L04, without a radial cut and a MED propagation model.
For all fluxes the spectral index is $\gamma_{\rm SNR}=2.4$ and $ E_{tot, {\rm SNR}} = 10^{49}$ erg.}
\label{fig:flusso}
\end{figure}

One of the components of our models is electrons produced by a smooth distribution of SNRs. Specifically, this component is from the SNRs located at  $R> R_{\rm cut}$ (where $ R_{\rm cut}$ can be allowed to go to zero).
The spatial distribution of these sources is taken from existing distribution models,  built from 
the catalogs of Galactic sources. Samples are usually corrected for observational selection effects, depending on the nature of the source data, for example radio or gamma rays. Most of SNR-based models  separate the vertical and the radial dependencies as:
\begin{equation}\label{eq:smoothdistr}
 \rho(r, z)= \rho_0 \,f(r) \, e^{-\frac{|z|}{z_0}}
\end{equation}
where $z_0=0.1$~kpc, $r$ is the distance from the Galactic center along the Galactic plane and $z$ indicates the location in the vertical (away from the plane) direction. In what follows, we fix the normalization coefficient $\rho_0 $ to 0.007 kpc$^{-3}$ such that the spatial distribution is normalized to unity within the diffusive halo. 
Our benchmark radial distribution model is G15, based on the `bright' sample of $69$ SNRs above the nominal surface brightness limit of $\Sigma_{1 \text{GHz}} = 10^{-20}$ W m $^{-2}$ Hz$^{-1}$ sr$^{-1}$, for which the Green's catalog of Galactic SNR is thought to be nearly complete. The G15 distribution can be parameterized as:
\begin{equation}
 f(r)_{G15}=  \left(\frac{r}{r_{\odot}}\right)^{a_1} e^{- a_2 \frac{r - r_{\odot}}{r_{\odot}}} 
  \label{eq:f(r)}
\end{equation}
where $a_1=1.09$ and $a_2=3.87$, the Galactocentric distance of the Earth is fixed to $r_{\odot} = 8.33$ kpc  \citep{Gillessen:2008qv}.
For comparison, we also consider the widely used radial distribution model L04 derived from the ATNF Galactic pulsar sample and given by:
\begin{equation}
f(r)_{\rm L04}= r^a  e^{{-\frac{r}{r_0}}}
\label{eq:L04}
\end{equation}
where $a=2.35$ and $r_0= 1.528$~kpc. 

In the semi analytical approach, steady-state solutions of Equation \eqref{eq:transport} can be solved by replacing the energy dependency $E$ with a pseudo-time $\tilde{t}(E)$. This leads to an inhomogeneous heat equation, whose solutions are given in terms of a Green's function formalism \cite{Baltz:1998xv}.  For a complete discussion of solutions and different approximations see \cite{Baltz:1998xv, 2010A&A...524A..51D} and \cite{2010pdmo.book..521S}.
The steady-state solution for the electron flux at Earth $\mathbf{x}_{\odot}= (r_{\odot}, 0, 0)$ is then the convolution:

\begin{equation}
\label{eq:fluxsolution}
\psi(\mathbf{x_{\odot}}, E)= \int dE_s \int  d^3 \mathbf{x_s} \, \mathcal{G}(\mathbf{x_\odot}, E \leftarrow \mathbf{x_s}, E_s) \, Q(E_s, \mathbf{x_s}),
\end{equation}
where the Green's function $\mathcal{G}(\mathbf{x_\odot}, E \leftarrow \mathbf{x_s}, E_s)$ represents the probability for an electron injected at $\mathbf{x_s}$ to reach the Earth with degraded energy $E<E_s$. 
The spatial integral is performed over the finite extent of the diffusive region. Hence, Green's functions have to account for boundary conditions. However, the radial boundary at $r_{\rm disc}=20$~kpc has been shown to be irrelevant at the Earth location when $r_{\rm disc}-r_{\odot}$ is of the same order, or larger than, $L$ \citep{2010A&A...524A..51D}.
In this case, the Green's function can be split  into radial and vertical term as $\mathcal{G}= (\mathcal{G}_r \times \mathcal{G}_z)/b(E) $, where $\mathbf{r}$ is the projection of the electron position in the $z=0$ plane. In what follows, we  will account for vertical boundary conditions only. Depending on the propagation scale, we will use the image method  \citep{Baltz:1998xv} or the Helmholtz eigenfunctions \citep{Lavalle:2006vb} to expand the vertical Green's functions $\mathcal{G}_z$ \citep{2010A&A...524A..51D}. 
 
Inserting the spatial distribution of SNRs from Equation \eqref{eq:smoothdistr} and the energy spectrum $Q(E)$ of Equation \eqref{eq:Q_SNR} in Equation \eqref{eq:fluxsolution}, the solution for the electron flux can be written as:
\begin{eqnarray}\label{eq:fluxsol}
\Phi(\mathbf{x}_\odot, E&) &= \frac{v}{4\pi} \frac{\rho_0  \Gamma_{*}}{b(E)} \int dE_s \, Q(E_s) \nonumber
\\
&& \int dx_s dy_s \,\mathcal{G}_r(x_\odot, y_\odot, E \leftarrow x_s, y_s, E_s ) \,f(r_s) \nonumber
\\
&& \int dz_s\, \mathcal{G}_z(z_\odot, E \leftarrow z_s, E_s) \, e^{-\frac{|z_s|}{z_0}} 
 \end{eqnarray}
where  $r_s=\sqrt{x_s^2 + y_s^2}$. As noted above, the Green's functions are taking into account the vertical boundary only. When the radial cut on the SNR position is applied, we implement it as a hollow cylindrical region around the Earth position in the source distribution, i.e., we set a hole in $\rho(r, z)$ defined by the condition $R \equiv |{\bf r} - {\bf r_{\odot}}|= R_{\rm cut}$.
Inside this hole, cataloged sources replace the smooth electron distribution from SNRs (and the resulting fluxes are obtained as discussed in the next subsection). 

In Figure~\ref{fig:flusso} we show an example of the electron flux that reaches the Earth, obtained from a smooth distribution of SNRs with an injection spectral index $\gamma_{\rm SNR}=2.4$ and $E_{tot, {\rm SNR}} = 10^{49}$ erg.  The results for both the MED and MAX propagation models are shown, as well as various choices of the cut-off value $ R_{\rm cut}$. We note that the reason to have a cut-off distance in the smooth distribution of SNRs is to allow us to introduce nearby discrete sources, as discussed in the next subsection.
Notice that the MED (blue lines) fluxes are higher than the MAX (red lines) in the no-cut (solid lines) and $ R_{\rm cut}=0.7$~kpc (dashed) cases, while the situation reverses for $R_{\rm cut}=3$~kpc (dot-dashed). This is because in the MAX propagation model the diffusion exponent $\delta$ is lower, and the half thickness of the diffusive halo $L$ is greater than in the MED case. For a small cut around the Earth this means that electrons diffuse more in the Galaxy, losing more energy. In contrast, when the value of $ R_{\rm cut}$ becomes comparable to the half-thickness of the diffusive halo,  electrons have less probability to reach the Earth. In this case, even if the diffusion exponent is higher, the MAX setup with $L\gg  R_{\rm cut}$ allows more electrons to reach us.  We note also that the L04 distribution (solid gray) predicts more electrons than the G15 model. This is because the L04 radial distribution predicts more sources in the solar circle. 

\subsection{Discrete distribution of sources  from catalogs}

Our model contains discrete sources, whose position and properties are taken from catalogs. 
Recently evidences for the existence of local CR sources ($<1$ kpc) has been found with the detection of Iron-60 isotop in CRs \citep{2016Sci...352..677B}.
We use catalogs both to specify the SNRs that are inside a cylinder around the Earth position of radius $ R_{\rm cut}$, for which we use the Green's catalog \citep{2015MNRAS.454.1517G}, and the
PWNe, which we take from pulsars in the ATNF catalog. 
In the two cases, the injection spectra are those defined in Equation (\ref{eq:Q_SNR}) for SNRs and in Equation (\ref{eq:Q_PWN}) for PWNe. Including sources from catalogs is especially important in electron and positron fluxes for energy greater than 100 GeV where local sources dominate \citep{2010A&A...524A..51D,DiMauro:2014iia,DiMauro:2015jxa,Manconi:2016byt}.

We solve the time-dependent diffusion equation in the point-source approximation. In this case, the propagation equation admits the analytical solution:
 \begin{equation}\label{eq:singlesourcesolution}
  \psi(\mathbf{x_{\odot}}, E,t) = \frac{b(E_s)}{b(E)} \frac{1}{(\pi \lambda^2)^{\frac{3}{2}}}  e^{ {-\frac{|\mathbf{x_{\odot}} -\mathbf{x_{s}} |^2}{ \lambda^2}} }Q(E_s)
\end{equation}
where $\mathbf{x_s}$ is the position of the source, and $E_s$ is the injection energy of electrons that cool down to $E$ because of energy losses $b(E)$ in a loss time:
\begin{equation}
 \Delta \tau (E, E_s) \equiv \int_E ^{E_s} \frac{dE'}{b(E')} = t-t_s
\label{eq:EsE}
\end{equation}
depending on the source age $t_s$. 
Therefore, the energy $E$ of an electron detected at Earth for a source with age $t_s$ that emits an electron with energy $E_s$ can be found from Equation (\ref{eq:EsE}).
The propagation scale $\lambda$ is defined as usual:
\begin{equation}\label{eq:lambda}
 \lambda^2= \lambda^2 (E, E_s) \equiv 4\int _E ^{E_s} dE' \frac{K(E')}{b(E')}\,. 
\end{equation}

\medskip
\section{Analysis and results}
\label{sec:results}

The features of the $e^+ + e^-$ spectrum are known to be dominated by the electron component produced by far SNRs at low energies, while at intermediate and high energies the components arising from local sources (either SNRs or PWNe) become important. Moreover, while nearby SNRs contribute only electrons, PWNe produce equal fluxes of positrons and electrons. Secondary positrons are comparable to PWN positrons below 10 GeV, although they never become a dominant component in the total $e^+ + e^-$ (except, maybe, at very high energies, beyond the cut-off energy of the source spectrum). For further details on these properties see, e.g.,~\cite{2010A&A...524A..51D,DiMauro:2014iia,DiMauro:2015jxa} and \cite{Manconi:2016byt} (where the analyses are performed in the same framework of Galactic transport we are using here), and references quoted therein. 

We investigate the role of the far and near SNR sources, the impact of PWNe on the high-energy tail of the {\it Fermi}-LAT spectrum, and we discuss whether a break in the injection spectrum or in the diffusion coefficient is required. The analysis is performed by fitting the new {\it Fermi}-LAT spectrum over their full energy range, by considering the whole set of leptonic contributions: primary electrons from SNRs, primary electrons and positrons from PWNe and secondary electrons and positrons. We use the {\it Fermi}-LAT $e^+ + e^-$ spectrum as reported in \cite{Abdollahi:2017nat}, and we consider the errors as given by the sum in quadrature of statistical and systematic uncertainties.

Since we are studying the total $e^+ + e^-$ spectrum it may happen that a good agreement with the data is found for a set of parameters that corresponds to a large positron flux, in excess of what is known from the PAMELA \citep{2013PhRvL.111h1102A} and AMS-02 \citep{2014PhRvL.113l1102A} measurements of this observable (and we will show in section \ref{sec:smooth} that this can indeed occur). In order to prevent this, we ``calibrate" our model by performing a fit to the AMS-02 positron-only flux, in order to determine priors on the parameters of the positron emission, that we then use in most of our analyses of the {\it Fermi}-LAT spectrum.


A summary of the different analyses that we perform with the corresponding free parameters and main hypotheses is presented in Table \ref{tab:analysis}.

\begin{table*}[t]
\center
\scalebox{0.99}{
\begin{tabular}{|c|cc|c|c|c|c|c|c|c|c|}
\hline 
Analysis & $  R_{\rm cut}$ [kpc] & $e^+$ priors & $ \gamma_{\rm{PWN}},\eta_{\rm{PWN}}$ & $\gamma_{\rm SNR}, E_{tot, \rm SNR}$& $ q$& $ B_{\rm Vela}$& $ \gamma_{\rm Vela}$& $ B_{\rm near}$& $ \gamma_{1,2 \rm SNR}$, $E_b^Q$\\
\hline
\rule{0pt}{3ex}
1& 0  & - & $\checkmark$ & $\checkmark$&  $ \checkmark$& n.a.& n.a.& n.a.& n.a. \\
\rule{0pt}{3ex} 
2& 0  & $\checkmark$& $\checkmark$ & $\checkmark$&  $ \checkmark$& n.a.& n.a.& n.a.& n.a. \\
\rule{0pt}{3ex}
3a& $0.7,3$ & $\checkmark$& $\checkmark$ & $\checkmark$&  $ \checkmark$& $\checkmark$& -& $ \checkmark$& n.a. \\
\rule{0pt}{3ex}
3b& $0.7$ & $\checkmark$& $\checkmark$ & $\checkmark$& $ \checkmark$& -& $\checkmark$& $ \checkmark$& n.a. \\
\rule{0pt}{3ex} 
4& 0 & $\checkmark$& $\checkmark$ & $\checkmark$&  $ \checkmark$& n.a.& n.a.& n.a.&$ \checkmark$  \\
\hline
\end{tabular}}
\caption{A summary of the main hypotheses and of the free parameters used in this work. For each analysis (1,2,3a,3b,4) we show the value for $R_{\rm cut}$ used for the SNR component and a check mark for the different parameters and priors that we use. For example, the priors from the AMS-02 positron spectrum have been used for all analysis except {\it Analysis-1}. 
The free parameters that are the spectral index ($\gamma_{\rm{PWN}}$) and  the efficiency ($\eta_{\rm{PWN}}$) for the PWNe ; the spectral index ($\gamma_{\rm SNR}$) and the normalization ($E_{tot, \rm SNR}$) for the smoothly distributed SNRs; the overall normalization for the secondary component ($q$), the value of the magnetic field ($B_{\rm Vela}$) and of the spectral index ($\gamma_{\rm Vela}$) for the Vela SNR, the value of the magnetic field for the near SNRs ($B_{\rm near}$), the parameters connected to the break in the spectral index for the SNR component ($ \gamma_{1,2 \rm SNR}$, $E_b^Q$). When the parameter is not applicable we indicate it as n.a. . }
\label{tab:analysis} 
\end{table*}

\subsection{Calibration of positron emission with {AMS-02} data}
\label{sec:calibrate}


We establish sensible values for the parameters that define the positron emission in our model by analyzing the AMS-02 positron flux \citep{2014PhRvL.113l1102A} at energies above 10 GeV. This is the same energy range for which the Fermi-LAT measured the $e^+ + e^-$ spectrum, and it is a choice that minimizes the impact of solar modulation on the determination of the model parameters. Notice that in all our analyses, solar modulation is included (a residual impact is also present at these high energies): we adopt a force-field approximation, and the Fisk potential is treated as a nuisance parameter.
In this analysis, the relevant free parameters are the efficiency $\eta_{\rm{PWN}}$ of the PWN for emission of positrons (see Equation \eqref{Wop}), the spectral index $\gamma_{\rm{PWN}}$ (see Equation \eqref{eq:Q_PWN}), and a normalization $q$ of the secondary positron emission.

\begin{figure*}[t]
\centering
\includegraphics[width=\columnwidth]{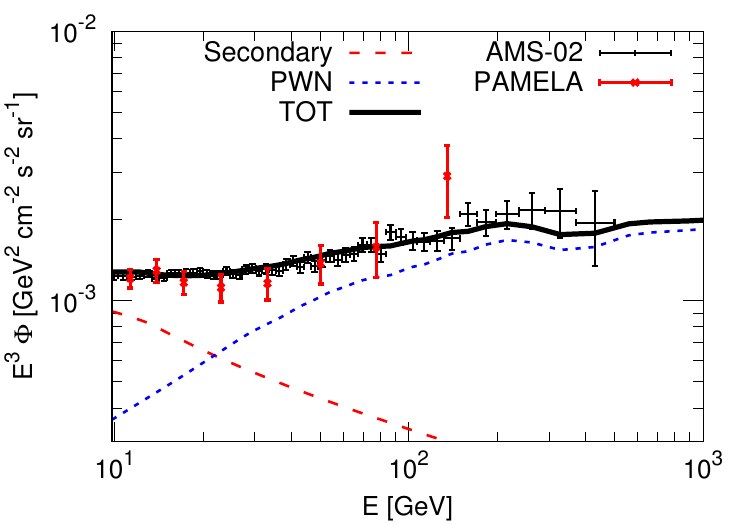}
\includegraphics[width=\columnwidth]{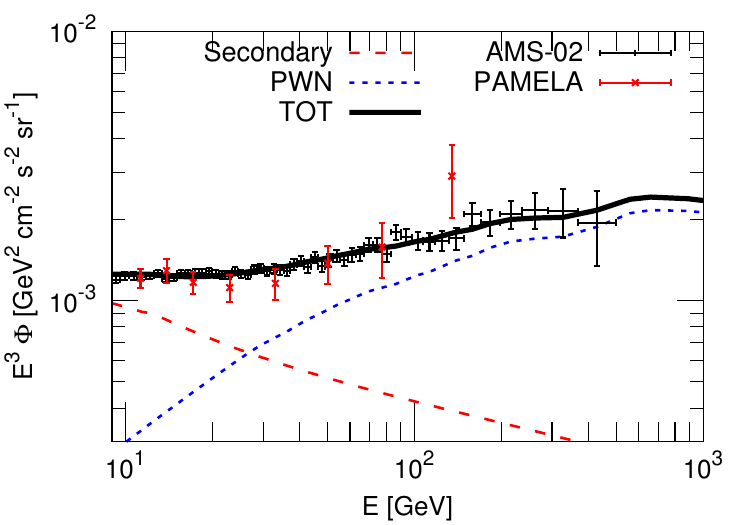}
\caption{Best fit to the AMS-02 positron flux, for energies greater than 10 GeV, for the MED (left panel) and MAX (right panel) propagation parameters. The black points are the AMS-02 data. The solid black line shows the best-fit model. The dotted blue line and the dashed red line show the PWN and secondary contributions. The red points are the PAMELA data for the  same observable.}
\label{fig:AMSfit}
\end{figure*}

Figure \ref{fig:AMSfit} shows the results for the best fit in the case of the MED (left panel) and MAX (right panel) transport parameters. Our model reproduces the AMS-02 positron spectrum, yielding a reduced chi-square $\chi_{\rm{red}}^2=\chi^2/{N_{\rm {d.o.f}}} = 0.51$ for the MED and $\chi_{\rm{red}}^2 = 0.61$ for the MAX propagation parameters.
The corresponding parameters are reported in Table \ref{tab:AMS}, while Table \ref{tab:priors} lists the ensuing priors that we will adopt in the rest of our analyses (they correspond to the $2\sigma$ intervals from the AMS-02 fit, and the priors are assumed to be flat in these intervals). We report in Table \ref{tab:AMS} and Table \ref{tab:priors}, respectively,  the best fit values and priors for secondary  and PWNe production for Kappl2015 and Genolini2015 propagation models that we will use in the rest of the paper to check how the choice of these more up-to-date propagation parameters affects our results. This allows us to evaluate the implications of the {\it Fermi}-LAT $e^++e^-$ spectrum, while remaining compatible with the AMS-02 (and PAMELA) measured positron flux.

\begin{table}[t]
\center
\begin{tabular}{|l|c|c|c|}
\hline
~ & $\eta_{\rm PWN}$ & $\gamma_{\rm PWN}$ & $q$ \\
\hline
\rule{0pt}{3ex}
MED & $0.0456^{+0.0012}_{-0.0011}$ & $1.80^{+0.04}_{-0.04}$ & $0.96^{+0.06}_{-0.06}$ \\
\rule{0pt}{3ex} 
MAX & $0.074^{+0.004}_{-0.003}$ & $1.90^{+0.04}_{-0.04}$ & $1.72^{+0.08}_{-0.08}$ \\
\rule{0pt}{3ex} 
Kappl2015 & $0.072^{+0.003}_{-0.002}$ & $1.91^{+0.04}_{-0.04}$ & $1.85^{+0.07}_{-0.08}$ \\
\rule{0pt}{3ex} 
Genolini2015 & $0.053^{+0.002}_{-0.002}$ & $1.90^{+0.04}_{-0.04}$ & $1.49^{+0.06}_{-0.06}$ \\
\hline
\end{tabular}
\caption{Best-fit parameters for the AMS-02 positron flux \citep{2014PhRvL.113l1102A}. The first/second/third/fourth row refers to the MED/MAX/Kappl2015/Genolini2015 CR propagation parameters.}
\label{tab:AMS}
\end{table}

\begin{table}[t]
\center
\begin{tabular}{|l|c|c|c|}
\hline
~ & $\eta_{\rm PWN}$ & $\gamma_{\rm PWN}$ & $q$ \\
\hline
\rule{0pt}{3ex}
MED & $[0.0437,0.0476]$ & $[1.72,1.88]$ & $[0.866,1.063]$ \\
\rule{0pt}{3ex} 
MAX & $[0.0693 ,0.0826]$ & $[1.83,1.97]$ & $[1.55,1.84]$ \\
\rule{0pt}{3ex} 
Kappl2015 & $[0.0672 ,0.0770]$ & $[1.83,1.99]$ & $[1.71,1.99]$ \\
\rule{0pt}{3ex} 
Genolini2015 & $[0.0493,0.0563]$ & $[1.82,1.98]$ & $[1.37,1.61]$ \\
\hline
\end{tabular}
\caption{Priors on the parameters that rule define positron emission from PWN and secondary production.}
\label{tab:priors}
\end{table}


\subsection{Smooth distribution of SNRs in the Galaxy}
\label{sec:smooth}

\begin{table*}[t]
\center
\begin{tabular}{|l|c|c|c|c|c|c|}
\hline
~ & $\eta_{\rm PWN}$&$\gamma_{\rm PWN}$&$E_{tot, \rm SNR}$ & $\gamma_{\rm SNR}$ & $q$ & $\chi_{\rm red}^2$ \\
\hline
MED & $0.059\pm0.009$ & $1.45\pm0.03$ & $5.67\substack{+0.3\\ -0.3} $ & $2.44\substack{+0.05 \\ -0.04} $ &$2.0$ & 0.68 \\ 
MAX & $0.049\pm0.003$ & $1.39\pm0.02$ & $12.5\substack{+0.2\\ -0.3} $ & $2.50$ &$2.0$ & $0.94$ \\ 
\hline
\end{tabular}
\caption{{\sl Analysis-1}. Best-fit parameters for the fit to the {\it Fermi}-LAT $e^++e^-$ spectrum for $q$ constrained to be in the range [0.5, 2.0] and $\gamma_{\rm{SNR}}<2.5$ in the case of MED and MAX propagation models and the G15 SNR distribution. $E_{tot, \rm SNR}$ is quoted in units of $10^{48}$ erg. The number of degrees of freedom is 38.}
\label{tab:SNRsmoothGreennopriors}
\end{table*}

\begin{figure*}[t]
\centering
\includegraphics[width=\columnwidth]{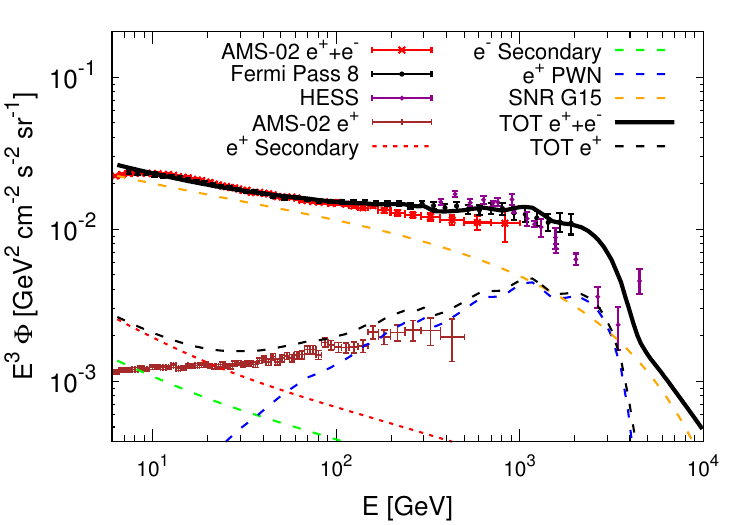}
\includegraphics[width=\columnwidth]{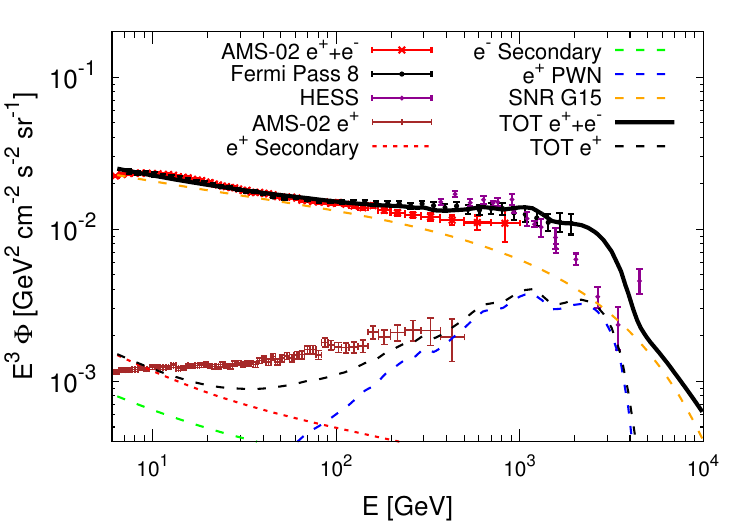}
\caption{{\sl Analysis-1.} Best fit to the {\it Fermi}-LAT $e^++e^-$ spectrum, obtained for the smooth SNR distribution with $ R_{\rm cut} = 0$ (i.e. no discrete local SNRs). Left (right) panels refer to the MED (MAX) cases. The fit assumes $q=[0.5,2.0]$ and $\gamma_{\rm{SNR}}<2.5$. The black points are the {\it Fermi}-LAT $e^++e^-$ spectrum. The black solid line shows the best-fit result. This is decomposed into the SNR electron contribution (orange dashed line), secondary electrons (green dashed line) and positrons (red dotted line) and positrons from PWNe (blue double-dot dashed line). For comparison, the plot also shows the AMS-02 positron flux (brown points), which can be compared with the total positron flux (dashed black line). The red and purple points are respectively the AMS-02 and H.E.S.S $e^++e^-$ spectrum.}
\label{fig:SNRsmoothnopriors}
\end{figure*}

\begin{figure*}[t]
\centering
\includegraphics[width=\columnwidth]{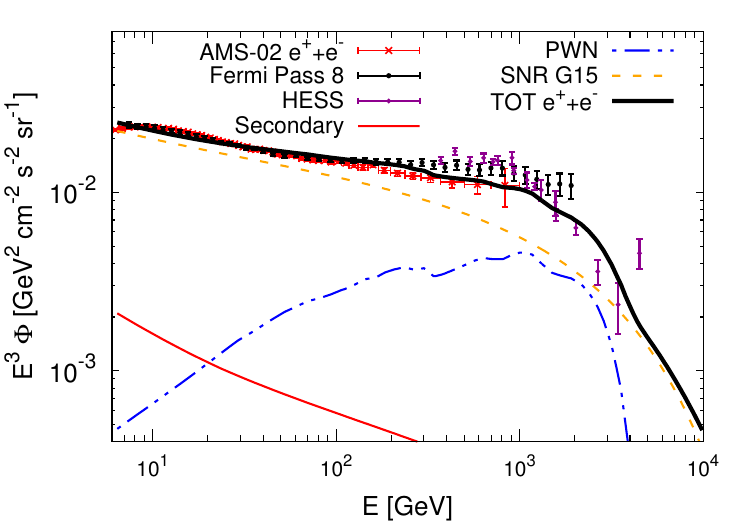}
\includegraphics[width=\columnwidth]{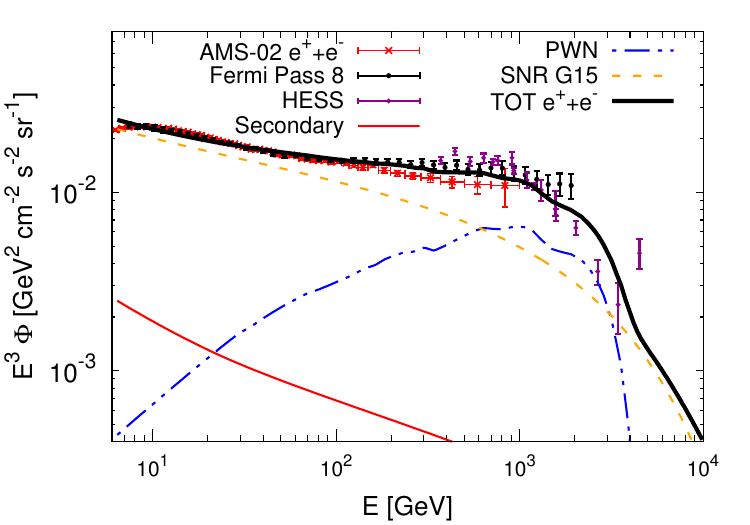}
\includegraphics[width=\columnwidth]{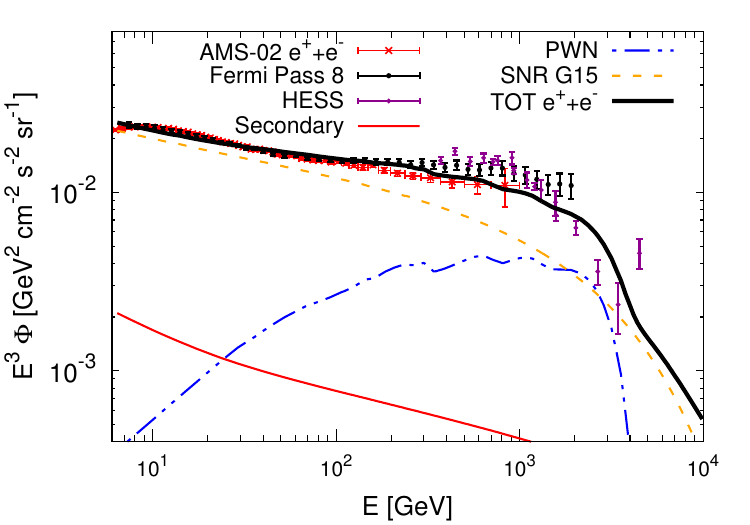}
\includegraphics[width=\columnwidth]{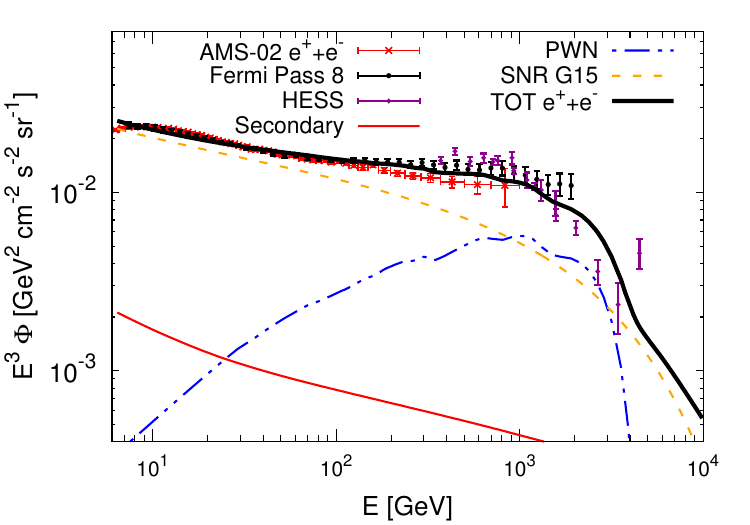}
\caption{{\sl Analysis-2}. Best fit to the {\it Fermi}-LAT $e^++e^-$ spectrum, obtained for the smooth SNR distribution with $ R_{\rm cut} = 0$ (i.e. no discrete local SNRs), assuming the priors informed by fitting the AMS-02 positron spectrum for the parameters that drive the positron contribution. The top left (top right) panel refers to the MED (MAX) while the bottom left (right) are for  Genolini2015 (Kappl2015) transport model, respectively. The $e^++e^-$ spectrum points are: {\it Fermi}-LAT (black), AMS-02 (red), H.E.S.S. (purple). The black line is the best fit to the {\it Fermi}-LAT spectrum. The orange dashed line shows the SNR electron contribution, the blue double-dot dashed line stands for the PWNe {$e^+ + e^-$} and the red solid line shows the secondary positrons.}
\label{fig:SNRsmoothGreen}
\end{figure*}

\begin{figure}[t]
\centering
\includegraphics[scale =0.25]{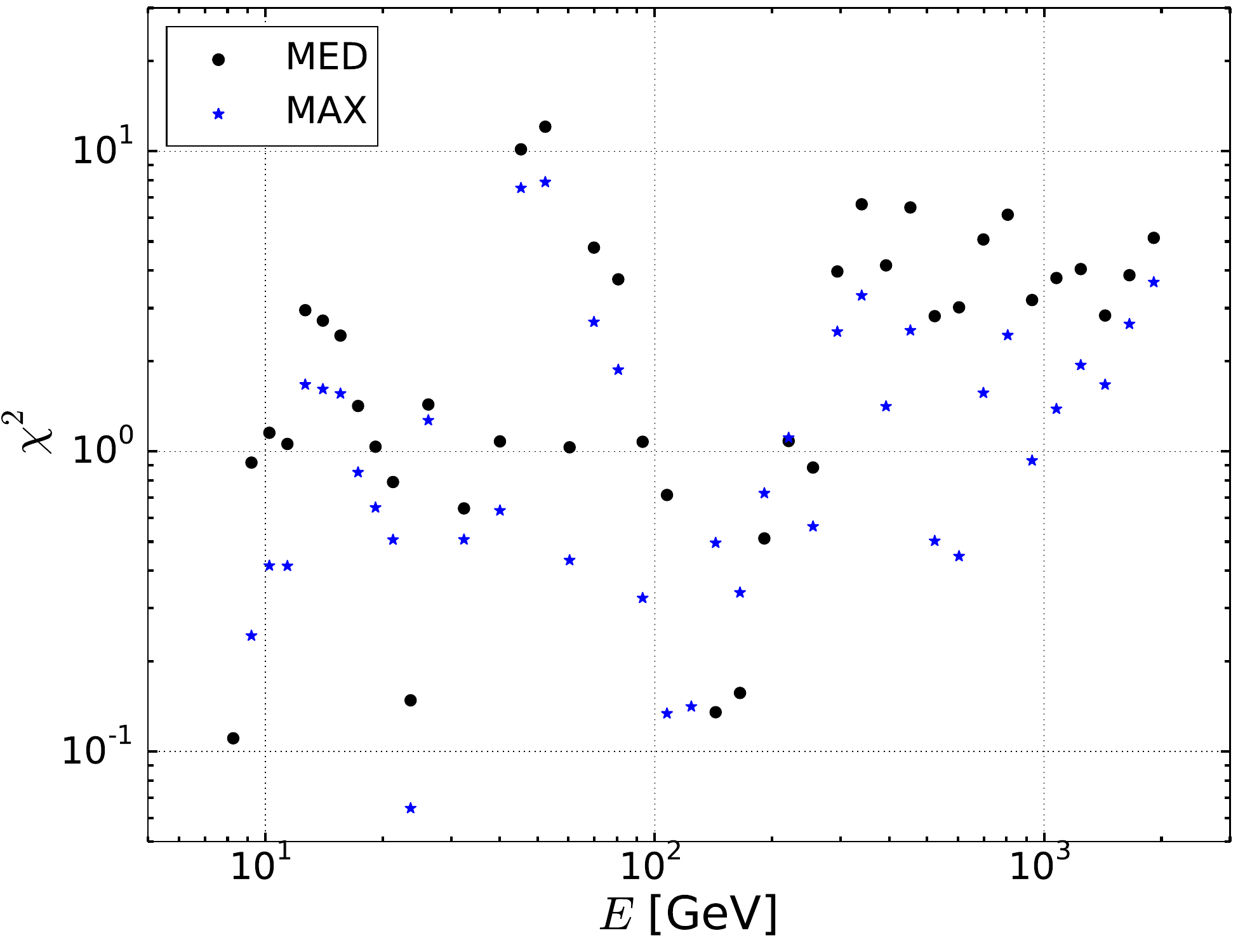}
\caption{{\sl Analysis-2.} Contribution to the $\chi^2$ in each {\it Fermi}-LAT energy bin, for the analyses reported in Figure \ref{fig:SNRsmoothGreen}.}
\label{fig:chi2}
\end{figure}


\begin{table*}[t]
\center
\begin{tabular}{|l|c|c|c|c|c|c|}
\hline
~ & $\eta_{\rm PWN}$&$\gamma_{\rm PWN}$& $E_{tot, \rm SNR}$  & $\gamma_{\rm SNR}$ & $q$ & $\chi_{\rm red}^2$ \\
\hline
MED & $0.0476$ & $1.72$ & $5.18\substack{+0.21\\ -0.20} $ & $2.410\substack{+0.009 \\ -0.009} $ &$1.06$ & 3.0 \\ 
\hline
MAX & $0.0826$ & $1.83$ & $14.0\substack{+0.6\\ -0.6} $ & $2.542\substack{+0.009 \\ -0.009} $ &$1.84$ & 1.6 \\ 
\hline
Kappl2015 & $0.0770$ & $1.83$ & $15.5\substack{+0.8\\ -0.8} $ & $2.560\substack{+0.010 \\ -0.010} $ &$1.99$ & 1.6 \\  
\hline
Genolini2015 & $0.0563$ & $1.82$ & $10.9\substack{+0.5\\ -0.5} $ & $2.532\substack{+0.010 \\ -0.010} $ &$1.61$ & 2.8 \\
\hline
\end{tabular}
\caption{{\sl Analysis-2.} Best-fit parameters for the fit to the {\it Fermi}-LAT $e^++e^-$ spectrum for the MED, MAX, Kappl2015 and Genolini2015 propagation models and a smooth G15 SNR distribution throughout the Galaxy. $E_{tot, \rm SNR}$ is quoted in units of $10^{48}$ erg. The number of degrees of freedom is 38.}
\label{tab:SNRsmoothGreen}
\end{table*}

The first analysis we perform on the Pass 8 {\it Fermi}-LAT $e^++e^-$ spectrum considers a model where SNRs are treated as a smooth population in the whole Galaxy, and for the moment we do not assume the AMS-02 priors of Table \ref{tab:priors} for the positron modeling. This is done in order to
investigate the direct implications of the  {\it Fermi}-LAT on the modeling of the cosmic leptonic components. For reference, we call this {\sl Analysis-1}.  In this case, the parameters that we leave free to vary in the fit are the PWN efficiency $\eta_{\rm PWN}$, the PWN index of the spectrum $\gamma_{\rm PWN}$, the normalization of the SNR spectrum $E_{tot, \rm SNR}$ (in units of $10^{48}$ erg), the SNR index of the spectrum  $\gamma_{\rm SNR}$ and the normalization factor $q$ of the secondary contribution. For this last quantity and for the SNR spectral index, we assume the following uniform priors: $q=[0.5,2.0]$ in order to allow some freedom for the calculated secondary positron spectrum and $\gamma_{\rm{SNR}}<2.5$, as expected for typical SNRs.

Results are shown in Figure \ref{fig:SNRsmoothnopriors} and Table \ref{tab:SNRsmoothGreennopriors}.  We obtain good agreement with the data both for the MED and MAX cases, with moderate and quite reasonable PWNe efficiencies around 5\%. We note that the model positron component is in agreement with the AMS-02 data, except in the MED case for energies below 30 GeV, a regime where solar modulation also might require a more refined analysis. These solutions, obtained by fitting the {\it Fermi}-LAT $e^+ +e^-$ spectrum alone without prior information on the positron contribution, are therefore quite satisfactory.
However, the normalization $q$ of the secondary production and the SNR spectral index lie mostly at the upper bounds of their priors, suggesting that if we allowed them to freely vary they would have unreasonable values. We have explicitly tried a fit without constraining their ranges, observing that in this case the secondary contribution is driven to be quite large of the order of 10: this has the consequence of greatly exceeding the AMS-02 measurements on positron spectrum. For this reason, from here on we consistently adopt throughout all our analyses the AMS-02 priors derived in the previous Section.
By looking at the best-fit configurations reported in Table~\ref{tab:SNRsmoothGreennopriors} (and this will be the case also for the results reported in the next sections) one can see that the best-fit values that we obtain for the parameter $E_{\rm tot, \rm SNR}$ are rather large. In fact, having $E_{\rm tot, \rm SNR} \approx 10^{49}$ erg implies that a fraction 10$^{-2}$ of the typical kinetic energy that is released in a supernova explosion is converted into $e^{\pm}$ pairs. This is in tension with typical values that are assumed for this fraction, which are around $10^{-5}$--$10^{-4}$ \citep[see the discussion in][]{2010A&A...524A..51D}. While an accurate study on this point would represent an important addition to our investigations, it is also important to point out that $E_{\rm tot, \rm SNR}$ is inversely proportional to the rate of supernova explosions $R$. In this work we are assuming $R = 1$, which is a rather low value if compared to the ones that are often quoted in literature (as an example in \cite{2010A&A...524A..51D} it  was assumed $R=4$). In addition, $E_{\rm tot, \rm SNR}$ strongly depends on the behavior of the spatial distribution of SNRs which, as discussed in \cite{2010A&A...524A..51D}, can exhibit large fluctuations (for example, by around a factor of 2 in the local neighbourhood). Lastly, as manifest from Table~\ref{tab:SNRsmoothGreen}, the best-fit value of $E_{\rm tot, \rm SNR}$ depends significantly on the Galactic propagation setup that is used. Such dependence could be even stronger if one would consider propagation models where the assumption of a uniform and isotropic diffusion is relaxed.

The results obtained by enforcing the AMS-02 priors are shown in Figure \ref{fig:SNRsmoothGreen}. 
The best-fit parameters are reported in Table~\ref{tab:SNRsmoothGreen}. We call this {\sl Analysis-2}.
First we note that the efficiency of PWNe lies at the upper bound of the priors and the PWNe index is close to its lower bound. This is because for energies around a few hundred GeV this model has a deficit with respect to the measurements; therefore the fit tends to fill this gap by increasing $\eta_{\rm{PWN}}$ and adopting the hardest $\gamma_{\rm{PWN}}$. Moreover, the spectral index of SNRs is 2.41 for MED and 2.54 for MAX, values consistent with the expectations for Fermi acceleration.

$\chi_{\rm{red}}^2$ is 3.0 for the MED and 1.6 for the MAX model and the energies where the fit does not provide a good representation of data are around $40-90$ GeV and for $E>250$ GeV. This can be seen in Figure \ref{fig:chi2}, where we break down the contributions to $\chi^2$ from the different energy bins. The MAX propagation model is significantly better than the MED model ($\Delta \chi^2=54$). However, Figure \ref{fig:chi2} shows that the MAX model also does not reproduce well the $e^++e^-$  in some energy ranges, especially at high energies. 

We repeated Analysis-2 using L04 instead of G15 for the SNR distribution (in the MED case) finding very similar results as those reported in Table \ref{tab:SNRsmoothGreen} for G15, except for the $E_{tot, {\rm SNR}}$ parameter which now is  $(4.49\pm 0.19)\times 10^{48}$ erg.
Indeed the spatial distribution in L04 does not change the spectral shape of the SNR contribution and only predicts a $15\%$ lower flux because the L04 density of SNRs is slightly greater than the G15 profile at the Earth position. As discussed in the Appendix \ref{ap:sp}, this picture does not seem to change if we include a spiral arm pattern in the distribution of SNRs to account for the presence of the Milky Way arms. In fact, when passing from the smooth distributions considered above to a more realistic one where sources are located along the arms, the electrons emitted by SNRs have to cover a larger distance before reaching Earth. As a consequence, their spectrum is softened by energy losses and therefore the tension with the potential hardening at high energies suggested by {\it Fermi}-LAT data becomes even stronger. 

We also tested the alternative propagation models Kappl2015 and Genolini2015. The best fit parameters that we find with these models are reported in Table~\ref{tab:SNRsmoothGreen} while in Fig.~\ref{fig:SNRsmoothGreen} we show the plot of the flux of the different components compared to {\it Fermi}-LAT data. The results we obtain with the Kappl2015 model are very similar to the ones we have with MAX model with the same $\chi^2$ and $\gamma_{\rm SNR}$ and with a slighlty greater $E_{tot, \rm SNR}$. On the other hand the Genolini2015 propagation setup gives a worse $\chi_{\rm red}^2$ of 2.8 with $\gamma_{\rm SNR}$ similar to the MAX and Kappl2015 cases.

In general, the model with a smooth distribution of SNRs appears inadequate especially above 50 GeV, where the {\it Fermi}-LAT data suggest a potential break. This might suggest that the {\it Fermi}-LAT spectral measurement requires a more detailed investigation of the mid/high energy range, where nearby SNRs (including the powerful Vela SNR) might have a role. 

\subsection{Electrons from far and near SNRs}

\begin{figure*}[t]
\centering
\includegraphics[width=\columnwidth]{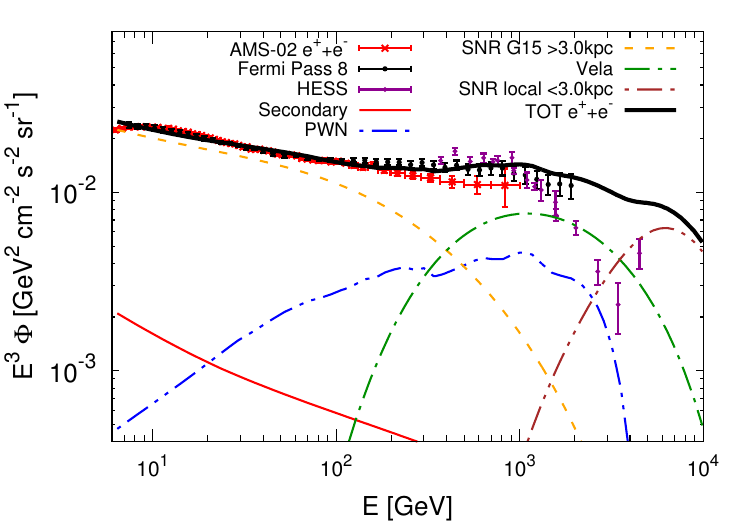}
\includegraphics[width=\columnwidth]{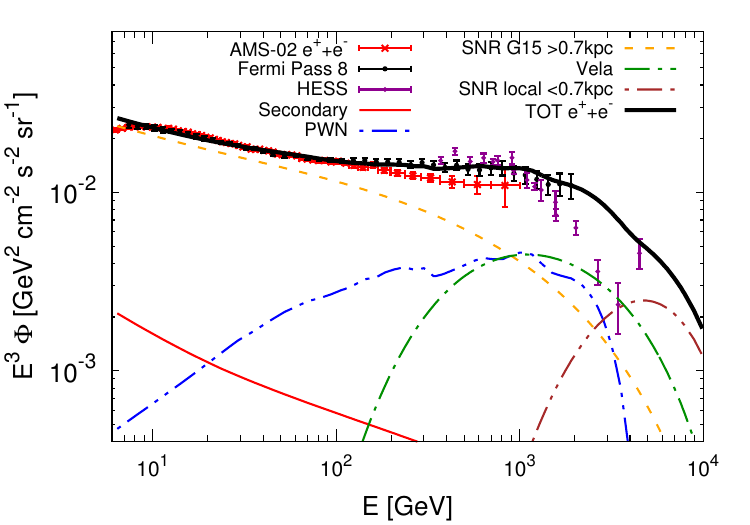}
\caption{{\sl Analysis-3a.} Best fit to the {\it Fermi}-LAT $e^++e^-$ spectral measurement, obtained for a SNR distribution composed of a near component ($R \leq  R_{\rm cut}$) and a far ($R >  R_{\rm cut}$) component, the latter being a smooth G15 distribution. The left panel refers to $ R_{\rm cut} = 3$ kpc, the right panel to $ R_{\rm cut} = 0.7$ kpc. In both cases, the propagation framework is MED. The $e^++e^-$ spectral points are {\it Fermi}-LAT (black), AMS-02 (red), H.E.S.S. (purple). The black line is the best fit to the {\it Fermi}-LAT spectral points. The orange dashed line shows the smooth SNR electron contribution, the blue double-dot dashed line stands for the PWNe {$e^+ + e^-$}  and the red solid line shows the secondary positrons. The green dot-dashed line shows the contribution of Vela and the purple double-dot-dashed line which emerges at the highest energies is the contribution of the near SNRs from Green's catalog.}
\label{fig:SNRsmoothVela}
\end{figure*}

\begin{figure*}[t]
\centering
\includegraphics[width=\columnwidth]{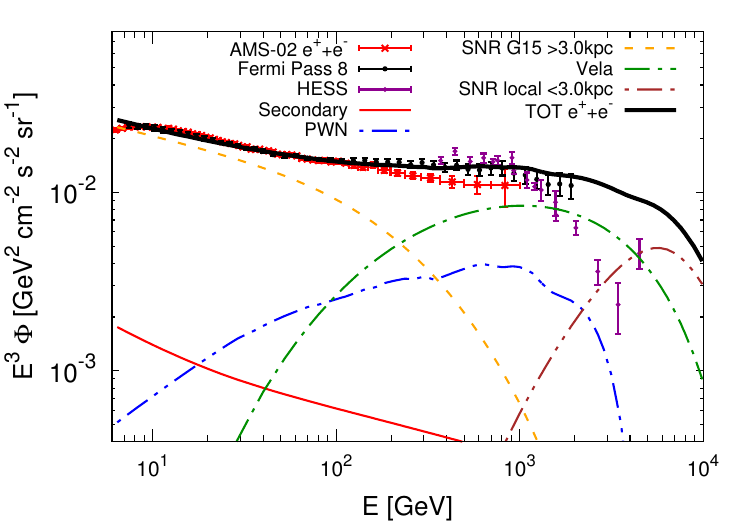}
\includegraphics[width=\columnwidth]{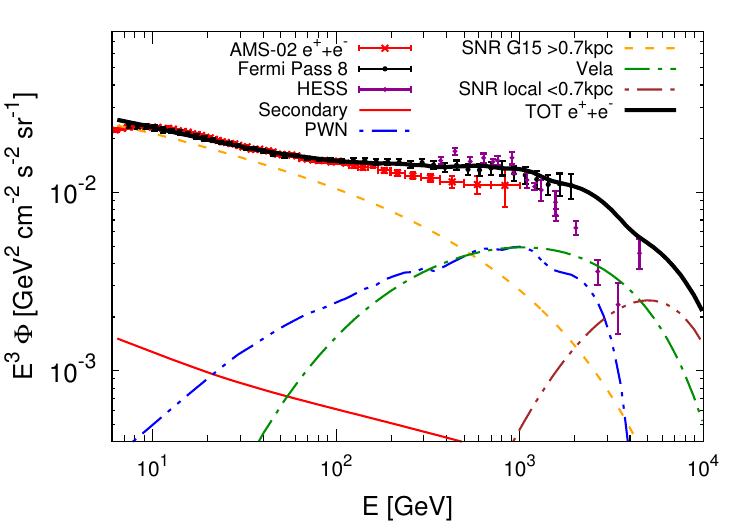}
\caption{{\sl Analysis-3a.} Same as in Figure \ref{fig:SNRsmoothVela}, for the MAX propagation model.}
\label{fig:SNRsmoothVelamax}
\end{figure*}

\begin{table*}[t]
\center
\begin{tabular}{|l|c|c|c|c|c|c|c|c|c|}
\hline
~ & $ R_{\rm cut}$ (kpc) & $\eta_{\rm PWN}$&$\gamma_{\rm PWN}$&$E_{tot, \rm SNR}$ & $\gamma_{\rm SNR}$ & $q$ & $B_{\rm Vela}$  & $B_{\rm near}$& $\chi_{\rm red}^2$ \\
\hline
MED & 3.0 & $0.0476$ & $1.72$ & $42.8\substack{+1.9\\ -1.8} $ & $2.144\substack{+0.019 \\ -0.019} $ &$1.06$ & $4.7 \substack{+0.2 \\ -0.2}$ & 20 & 2.0 \\ 
\hline
MED & 0.7 &$0.0476$ & $1.72$ & $9.4\substack{+0.7\\ -0.6} $ & $2.392\substack{+0.006 \\ -0.005} $ &$1.06$ & $6.3 \substack{+0.3 \\ -0.3}$ & 20 & 0.75 \\ 
\hline
\hline
MAX & 3.0  &$0.0826$ & $1.97$ & $25.0\substack{+0.3\\ -0.2} $ & $2.244\substack{+0.003 \\ -0.002} $ &$1.84$ & $4.0 \substack{+0.2 \\ -0.2}$ & 20 & 0.67 \\ 
\hline
MAX & 0.7 & $0.0693$ & $1.83$ & $23.6\substack{+0.3\\ -0.2} $ & $2.563\substack{+0.002 \\ -0.002} $ &$1.55$ & $5.7 \substack{+0.3 \\ -0.3}$ & 20 & 0.39 \\
\hline
\end{tabular}
\caption{{\sl Analysis-3a.} Best-fit parameters for the fit to the {\it Fermi}-LAT $e^++e^-$ spectral measurement for the MED propagation model, when the SNR distribution is separated into a far component ($R> R_{\rm cut}$) for which the smooth source distribution is G15, and a near component where the contribution from the individual SNRs of Green's catalog with a distance less than $ R_{\rm cut}$ are added. $E_{tot, \rm SNR}$ is quoted in units of $10^{48}$ erg, while magnetic field intensities are in $\mu$G. The number of degrees of freedom is 38.}
\label{tab:SNRfarnear}
\end{table*}

We found in the previous section that a smoothly distributed population of SNRs is not able to provide a good fit to the $e^++e^-$ spectrum over the entire energy range measured by {\it Fermi}-LAT,
once constraints on parameter ranges derived from the AMS-02 positron spectrum are taken into account. 
Therefore, we now allow for more freedom in our treatment of the SNR contribution, by considering far and near SNRs as separate kinds of sources in our fitting procedure. As detailed in Section \ref{sec:contributions}, this is realized by setting the parameter $R_{\rm cut}$ to values different from zero. The properties of the local SNRs are taken from  \cite{Green:2014cea}.
Green's catalog includes only a few sources able to shape the high energy flux, with the Vela SNR in a dominant position.
The normalization of the injection spectrum $Q_{0,{\rm SNR}}$ (see Equation \eqref{eq:Q_SNR}) for the Vela SNR can be related to the synchrotron emission of the electrons propagating in the magnetic field:  
\begin{eqnarray}
\label{eq:vela_norm}
Q_{0, \rm{Vela}} &=  &\left(\frac{1.2 \cdot 10^{47}}{\rm{GeV}}\right) \cdot (0.79)^{\gamma_{\rm{Vela}}} \left( \frac{d_{\rm{Vela}}}{\rm{kpc}} \right)^2 
 \\
&&\left(\frac{\nu}{\rm GHz} \right)^{(\gamma_{\rm{Vela}}-1)/2} 
\left( \frac{B_{\rm{Vela}}}{100\mu\rm{G}} \right)^{-(\gamma_{\rm{Vela}}+1)/2} \left( \frac{B^{\nu}_r}{\rm{Jy}} \right), \nonumber
\end{eqnarray}
where $d_{\rm{Vela}}$ is the distance to Vela, which we assume to be $d_{\rm{Vela}} = 0.293^{+0.019}_{-0.017}$ kpc \citep{2003ApJ...596.1137D}, and $B^{\nu}_r$ is the differential intensity measured at radio frequency $\nu$. The spectral index $\gamma_{\rm{Vela}}$ can be written in terms of the index of the synchrotron emission $\gamma_{\rm{Vela}} = 2\alpha_r+1$. 

Early observations of Vela \citep{1958AuJPh..11..550R} detected three regions of intense radio emission: Vela X, interpreted as the radio source associated with the Vela PWN, Vela Y, and Vela Z, which, because of their steeper radio spectrum than that of Vela X, are assumed to be part of the shell-type SNR.

As shown by \cite{2001A&A...372..636A} the emission from Vela Y and Z has a radio spectral index of $\alpha_r=0.70 \pm 0.10$ and $\alpha_r=0.81 \pm 0.16$, respectively, while the radio fluxes at 960 MHz are $(588\pm72)$ Jy and $(547\pm83)$ Jy. We consider for the radio flux the sum of the fluxes from Vela Y and Z, $B^{\nu}_r = (1135\pm110)$ Jy, and for the index the average of the spectral indices of Vela Y and Z $\gamma_{\rm{Vela}} = 2.50\pm0.30$ (since they are very similar).
We apply the same Equation (\ref{eq:vela_norm}) to the Cygnus Loop and  the other near SNRs. For those sources we take the parameters from Green's catalog.

We leave free the normalizations of the fluxes emitted by the two most powerful local SNRs, Vela and the Cygnus Loop. A change in the normalization of the flux can be interpreted as  a change in the magnetic field of the remnant: $Q_{\rm 0, SNR}\propto (B/100\rm{\mu G})^{-(\gamma_{\rm SNR} +1)/2}$, where  $B$ is the intensity of the magnetic field. We start by assuming the two magnetic fields to be in the range $10< (B_{\rm{Vela}}/\muG) <200$ and $20< (B_{\rm{near}}/\muG) <60$ for the magnetic field of Vela and the Cygnus Loop \citep{0004-637X-741-1-44}, respectively. For definiteness, we also take the magnetic fields of all the other local SNRs to be equal to the magnetic field of the Cygnus Loop.

The free parameters in this analysis are $\eta_{\rm PWN}$, $\gamma_{\rm PWN}$, $E_{tot, \rm SNR}$, $\gamma_{\rm SNR}$,  $q$, $B_{\rm{Vela}}$ and $B_{\rm{near}}$, while $\gamma_{\rm Vela}$ is fixed to 2.5 ({\sl Analysis-3a}). The best-fit configurations for two different values of the parameter $R_{\rm cut}= 0.7$ kpc and 3 kpc are shown in Figure \ref{fig:SNRsmoothVela} for MED and in Figure \ref{fig:SNRsmoothVelamax} for MAX. The best-fit parameters are reported in Table~\ref{tab:SNRfarnear}. For $R_{\rm{cut}} = 0.7$ kpc, the agreement with data is remarkably good, both in the MED and MAX cases. On the other hand, setting $R_{\rm{cut}}$ to 3 kpc gives a much worse fit in the MED case, while it is still quite good for the MAX case. The situation for the MED propagation setup and $R_{\rm{cut}}= 3$  kpc can be seen in the left panel of Figure \ref{fig:SNRsmoothVela}; in this case the model under-predicts the data around a few hundred GeV. This is probably because Green's catalog of SNRs, from which we select the local sources, contains only the nearest and brightest objects that contribute to $E>1$ TeV. The catalog is probably incomplete for those sources that are older and fainter and which should contribute at a few hundred GeV. Setting the radial cut at 0.7 kpc alleviates the tension with the data because all the fainter and older sources are incorporated into the smooth distribution of SNRs and the nearby component is dominated by Vela and the Cygnus Loop SNRs, which we are including explicitly in the model.

\begin{figure*}[t]
\centering
\includegraphics[width=\columnwidth]{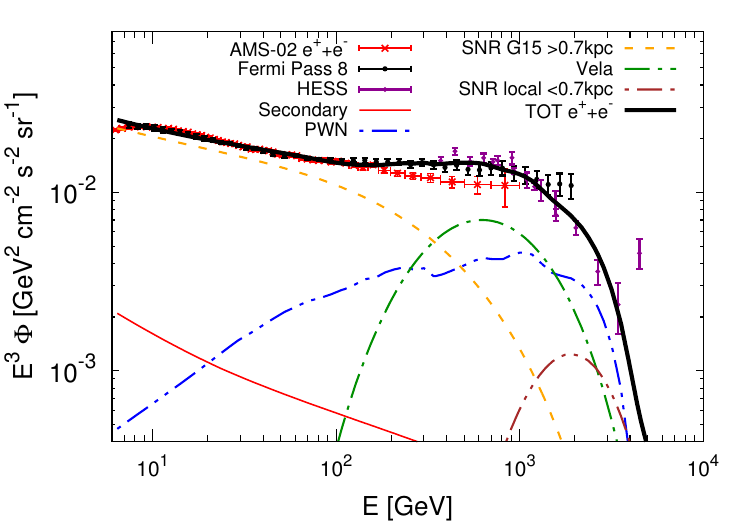}
\includegraphics[width=\columnwidth]{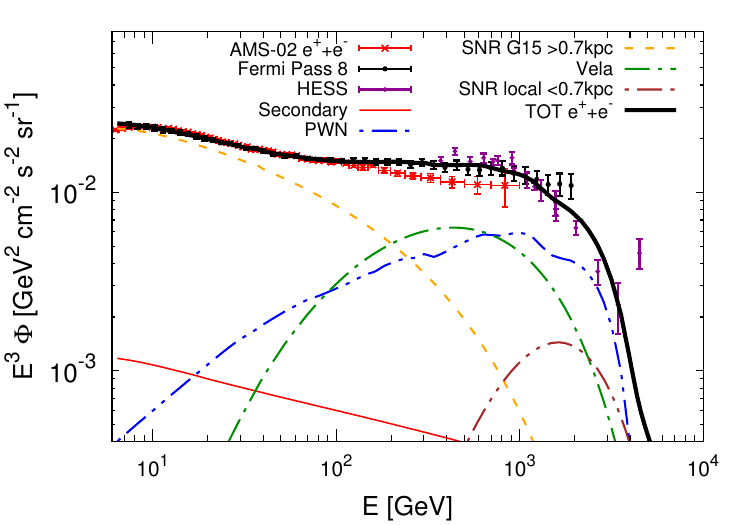}
\caption{{\sl Analysis-3a.} Same as in Figure \ref{fig:SNRsmoothVela} for the MAX  and MED propagation model with $E_c=1$ TeV.}
\label{fig:SNRsmoothVelacut}
\end{figure*}

We also test values for SNR cutoff energies different from 5 TeV, the benchmark value in our analysis.
The values $E_c = [0.6, 1.0, 2.0, 3.0]$ TeV are also used with the MAX and MED propagation parameters with the result that the goodness of fit and the values of the best-fit parameters are consistent with those found for 5 TeV (reported in Table~\ref{tab:SNRfarnear}) if $E_c>2$ TeV. On the other hand for $E_c<2$ TeV the fit worsens significantly because the nearby SNRs, mainly Vela and the Cygnus Loop, do not explain the highest-energy spectral points due to the low energy of the cutoff. In Figure \ref{fig:SNRsmoothVelacut} we show the result of a fit as in Figure \ref{fig:SNRsmoothVelamax} but with $E_c=1$ TeV for the MED (right panel) and MAX propagation parameters (left panel).
It is evident from these two plots that setting the cutoff energy of the SNR emission at 1 TeV results in a sizeable reduction of the flux from local SNRs such as the Cygnus Loop. This happens because these sources have ages and distances for which the peak of their fluxes is expected to be at higher energies (around 5 TeV), as shown in Figure \ref{fig:SNRsmoothVelamax}.

In the fits, the magnetic field of the Cygnus Loop takes the lowest value allowed by the prior ($B_{\rm{near}} = 20\,\mu$G) while $B_{\rm{Vela}}$ is found to be in the range $(5-6)\,\mu$G when $R_{\rm cut}$ = 0.7 kpc.
The magnetic field of the Vela SNR is significantly smaller than the value derived in \cite{Sushch:2013tna}. In that paper, based on the modeling of the synchrotron emission from Vela using an advanced hydrodynamical framework, the magnetic field of Vela Y and Z is found to be $46\,\mu$G 
and $30\,\mu$G, respectively. In \cite{Sushch:2013tna} the index of the injection spectrum has been derived to be $\gamma_{\rm Vela}=2.47\pm0.09$ and 
the total energy emitted by Vela in the form of electrons to be $E_{tot, \rm{Vela}} = (2.4\pm0.2)\cdot10^{47}$ erg. 
This last quantity is directly related to our modeling of Vela, and in particular to $Q_{\rm 0, Vela}$, by Equation \eqref{eq:eleenergy}. 
%
We therefore try a final fit to {\it Fermi}-LAT spectrum ({\it Analysis-3b}) by fixing the Vela magnetic field to $B_{\rm{Vela}}=38\,\mu$G, 
which is the average of the magnetic fields of Vela Y and Vela Z, and vary $Q_{\rm 0, Vela} $  and $\gamma_{\rm Vela}$ within the $2\sigma$ intervals of these parameters as given by \cite{Sushch:2013tna}.

\begin{figure*}[t]
\centering
\includegraphics[width=\columnwidth]{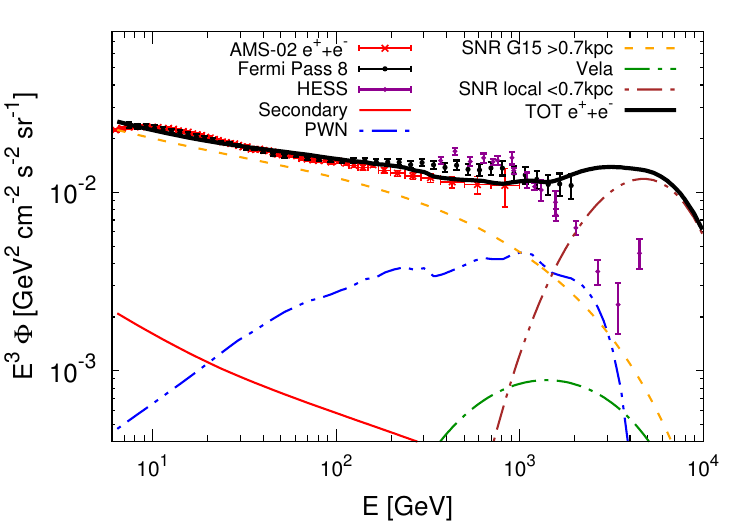}
\includegraphics[width=\columnwidth]{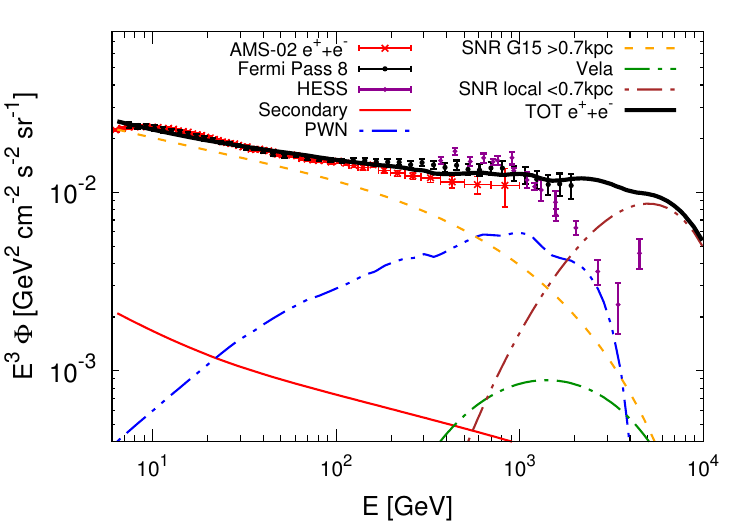}
\caption{{\sl Analysis-3b.} As in Figure \ref{fig:SNRsmoothVela} and Figure \ref{fig:SNRsmoothVelamax}, but for the MED (left panel) and MAX (right panel) propagation parameters and using as priors on $Q_0$ of Vela and $\gamma_{\rm Vela}$ the $2\sigma$ intervals of the values derived in  \cite{Sushch:2013tna}.}
\label{fig:smoothSNRVeladef}
\end{figure*}

The results of this fit are shown in Figure \ref{fig:smoothSNRVeladef} for the MED (left panel) and MAX (right panel) propagation parameters. The best-fit parameters are in Table~\ref{tab:smoothSNRVeladef}.  The ${\chi}^2_{\rm red}$ values with either MED or MAX propagation parameters are much worse than in the previous case, for which the Vela SNR parameters were free to vary. This is primarily because fixing the parameters $B_{\rm{Vela}}$, $E_{tot, \rm{Vela}}$ and $\gamma_{\rm Vela}$ fixed to the values derived in \cite{Sushch:2013tna} implies an electron flux much smaller than obtained 
with the Vela SNR parameters specified in Table~\ref{tab:SNRfarnear}. 
This value of $B_{\rm{Vela}}$ makes the spectrum lower by about one order of magnitude, creating a deficit of electrons around a few hundred GeV.
Indeed, making the same fit but without considering any prior on $E_{tot, \rm{Vela}}$ we find $E_{tot, \rm{Vela}}=32\cdot 10^{48}$ erg, which is more than an order of magnitude larger than in \cite{Sushch:2013tna}.

\begin{table*}[t]
\center
\begin{tabular}{|l|c|c|c|c|c|c|c|c|}
\hline
~ &  $\eta_{\rm PWN}$&$\gamma_{\rm PWN}$&$E_{tot, \rm SNR}$ & $\gamma_{\rm SNR}$ & $q$ & $\gamma_{\rm Vela}$ & $B_{\rm near}$ & $\chi_{\rm red}^2$ \\
\hline
MED & $0.0476$ & $1.72$ & $8.26\substack{+0.45\\ -0.40} $ & $2.358\substack{+0.009 \\ -0.008} $ &$1.06$ & $2.29$ & $43\pm3$ & 2.6 \\ 
\hline
MAX & $0.0830$ & $1.83$ & $14.7\substack{+0.8\\ -0.7}$ & $2.462\substack{+0.011 \\ -0.010}$  &$1.84$ &  $2.29$ & $53\pm4$& 1.52 \\ 
\hline
\end{tabular}
\caption{{\sl Analysis-3b.} Best-fit parameters for the fit to {\it Fermi}-LAT $e^++e^-$ spectral measurement in the case of MED (top panel) or MAX (bottom) propagation model with SNRs divided into a smooth component for objects with $R>0.7$ kpc and near sources taken from Green's catalog. 
$E_{tot, \rm SNR}$ is quoted in units of $10^{48}$ erg and the magnetic field $B_{\rm near}$ in $\mu G$. The number of degrees of freedom is 38.}
\label{tab:smoothSNRVeladef}
\end{table*}

\medskip
\section{Interpretations with a break in the injection spectrum or in the diffusion coefficient}
\label{sec:breaks}

In the previous sections we interpreted the {\it Fermi}-LAT $e^+ + e^-$ spectrum by using different models for the spatial distribution of electron and positron sources. In this section we study whether the apparent hardening around 50 GeV could be explained by a break in the injection spectrum or in the diffusion coefficient. In this analysis, we will use the G15 smooth SNR distributions for the whole Galaxy, without considering a separate near component ($R_{\rm cut}$=0).

\begin{figure*}[t]
\centering
\includegraphics[width=\columnwidth]{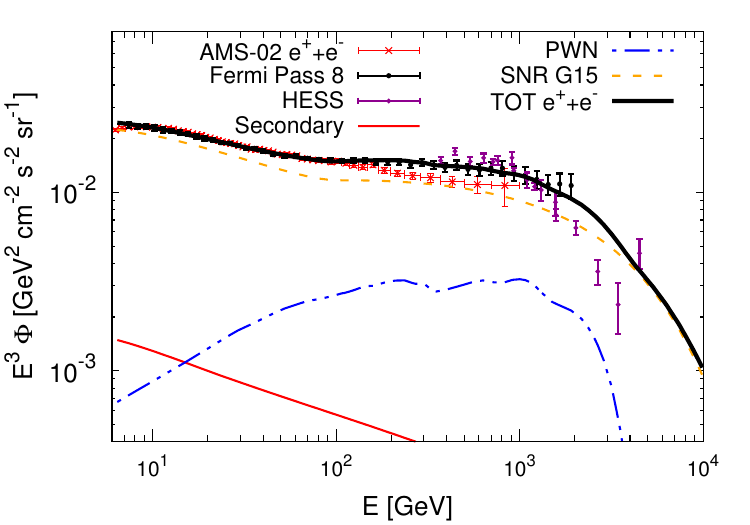}
\includegraphics[width=\columnwidth]{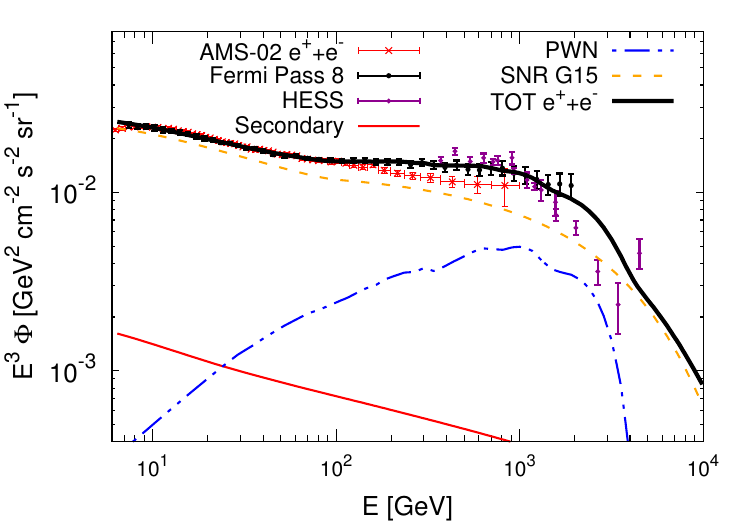}
\includegraphics[width=\columnwidth]{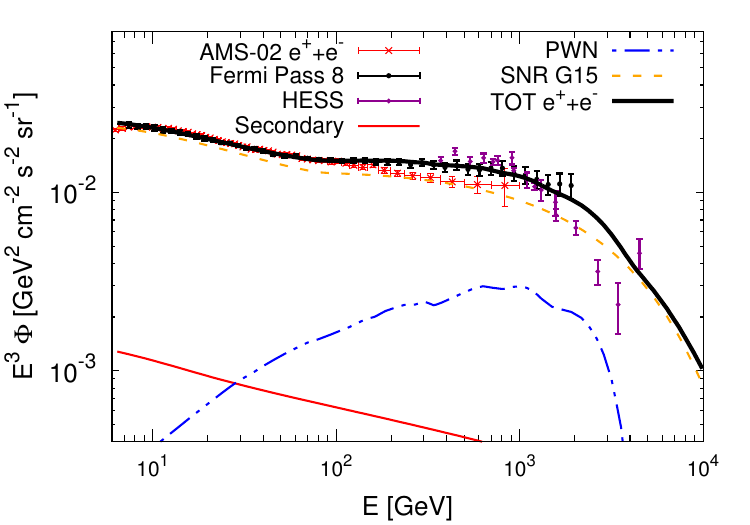}
\includegraphics[width=\columnwidth]{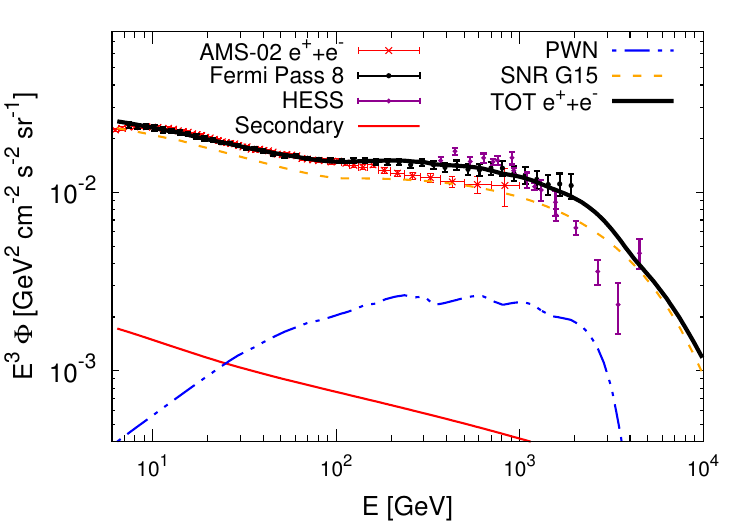}
\caption{{\sl Analysis-4.} Best-fit to the {\it Fermi}-LAT $e^++e^-$ spectral measurement, obtained from a SNR smooth population  with a break in the injection spectrum,
using the MED (top left panel), the MAX (top right panel), Kappl2015 (bottom left panel) and Genolini2015 propagation model (bottom right panel).}
\label{fig:SNRbreak}
\end{figure*}

In order to account for a break related to the injection of electrons, the SNR spectrum is now modeled as a broken power law:
\begin{eqnarray}
\label{eq:broken_Q}
Q(E) =
\left\{
\begin{array}{rl}
& Q_{0, \rm SNR} \left( \frac{E}{E_0}  \right)^{-\gamma_{1,\rm{SNR}}} 
\phantom{ \hspace{-1cm} \left( \frac{E_{\rm b}^Q}{E_0}  \right)^{\gamma_{2,\rm{SNR}}-\gamma_{1,\rm{SNR}}} }
 E \leq E_{\rm b}^Q, \\
& Q_{0, \rm SNR}\left( \frac{E}{E_0}  \right)^{-\gamma_{2,\rm{SNR}}} \left( \frac{E_{\rm b}^Q}{E_0}  \right)^{\Delta \gamma_{\rm SNR} }   E > E_{\rm b}^Q,
\end{array}
\right.
\end{eqnarray}
where $\Delta \gamma_{\rm SNR} = \gamma_{2,\rm{SNR}}-\gamma_{1,\rm{SNR}} $.
The free parameters of our model are now 
$\eta_{\rm PWN}$, $\gamma_{\rm PWN}$, q, $E_{tot, \rm SNR}$, $\gamma_{1, \rm SNR}$,  $\gamma_{2, \rm SNR}$ and $E_{\rm b}^Q$ .
The best-fit model for this case ({\sl Analysis-4}) is shown in  Figure \ref{fig:SNRbreak} and the corresponding parameters are listed in Table~\ref{tab:SNRsmoothbreak}. We find that this option reproduces the spectral measurement very well, making the possibility of a broken power law for the injection spectrum of SNRs viable. 
For both the MED, MAX, Kappl2015 and Genolini2015 models, the implications are that the break in the injection spectrum would occur at an energy of $E_{\rm b}^Q$=100 GeV, larger than the effective energy of the break in $e^+ + e^-$ spectrum. 
 This difference between the position of the break at injection and at the Earth can be due to the propagation history. 
 Electrons diffuse and cool radiatively from the sources to the Earth. 
 The change in the spectral index is $\Delta\gamma_{\rm SNR} = -0.42 \pm0.02$. This spectral 
 hardening could be due to the physics of the SNR shocks \citep{Caprioli:2010ne} or 
 to an emerging SNR population with a harder injection index. 

\begin{table*}[t]
\center
\begin{tabular}{|l|c|c|c|c|c|c|c|c|}
\hline
~ &  $\eta_{\rm PWN}$&$\gamma_{\rm PWN}$&$E_{tot, \rm SNR}$ & $\gamma_{1, \rm SNR}$ & $\gamma_{2, \rm SNR}$& $E_{\rm b}^Q$ & q & $\chi_{\rm red}^2$ \\
\hline
MED & $0.0476$ & $1.72$ & $12.5\substack{+0.9\\ -0.8} $ & $2.608\substack{+0.011 \\ -0.010} $ & $2.185\substack{+0.018 \\ -0.016} $ & $100\substack{+15 \\ -15} $ & 1.063 & 0.28 \\ 
MAX & $0.0693$ & $1.83$ & $26.6\substack{+0.4\\ -0.4} $ & $2.673\substack{+0.008 \\ -0.007} $ & $2.378\substack{+0.017 \\ -0.016} $ & $100\substack{+15 \\ -15} $ & 1.84 & 0.24 \\ 
Kappl2015 & $0.0672$ & $1.91$ & $37.1\substack{+0.5\\ -0.5} $ & $2.744\substack{+0.011 \\ -0.010} $ & $2.410\substack{+0.019 \\ -0.018} $ & $95\substack{+15 \\ -15} $ & 1.99 & 0.25 \\ 
Genolini2015 & $0.0493$ & $1.97$ & $27.2\substack{+0.2\\ -0.2} $ & $2.665\substack{+0.008 \\ -0.007} $ & $2.410\substack{+0.022 \\ -0.022} $ & $105\substack{+15 \\ -15} $ & 1.61 & 0.28 \\ 
\hline
\end{tabular}
\caption{{\sl Analysis-4.} Best fit to the {\it Fermi}-LAT $e^++e^-$ spectral measurement, obtained for Best-fit parameters  in the case of the MED (top) and MAX (bottom) propagation model, obtained with SNRs with a break in the injection spectrum. $E_{tot, \rm SNR}$ is quoted in units of $10^{48}$ erg. The number of degrees of freedom is 38.}
\label{tab:SNRsmoothbreak}
\end{table*}

\begin{figure}[t]
\centering
\includegraphics[width=\columnwidth]{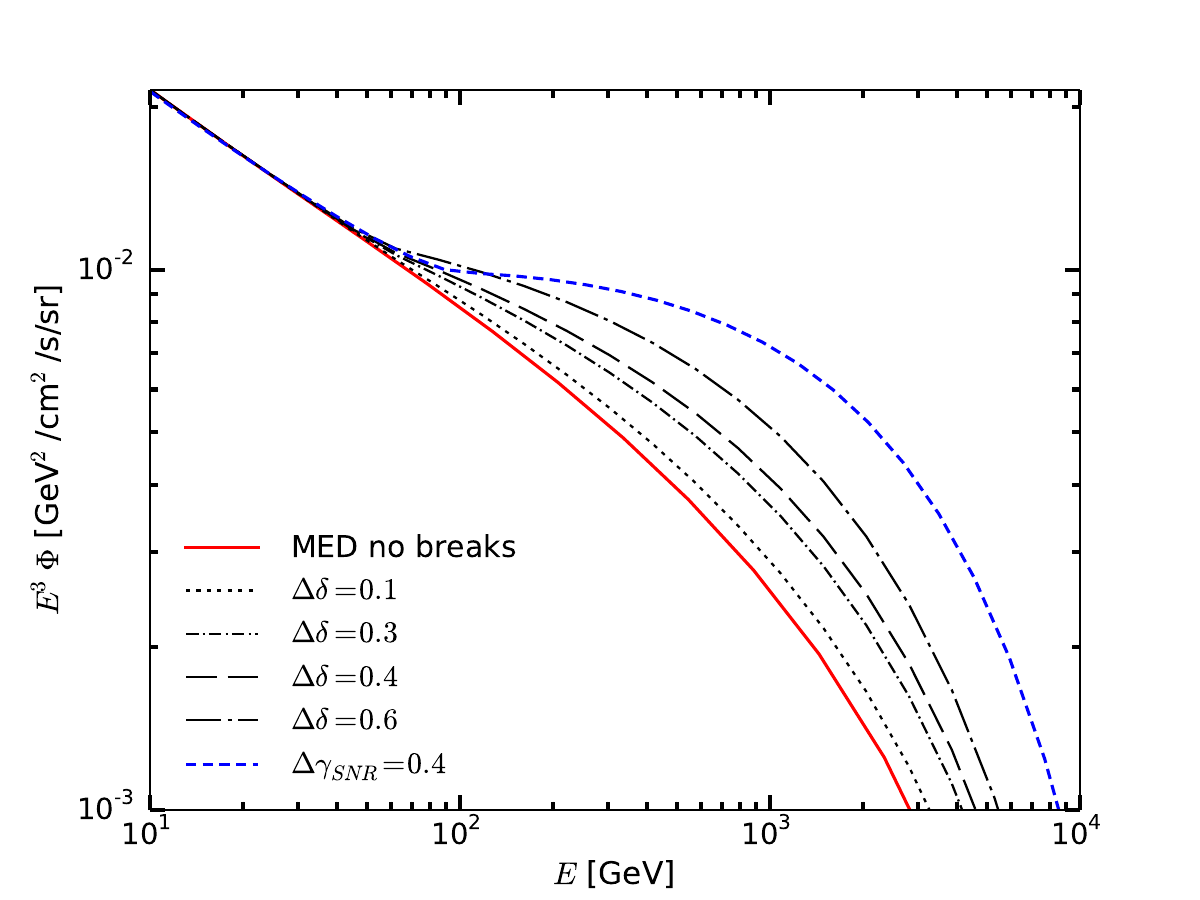}
\caption{Electron flux from the G15 smooth population distribution for SNRs with a break in the diffusion coefficient at $E_{\rm b}^K=60$ GeV (black lines) compared with the standard case with no breaks (red lines), and with the case of a break in the injection spectrum at $E_{\rm b}^Q=100$ GeV (blue line). Propagation is computed using the MED parameters.}
\label{fig:break_K}
\end{figure}

A change in the spectral shape of electron and positron fluxes could also be due to a spectral break in the diffusion coefficient $K(E)$. 
Such an effect has been proposed to account for the hardening in the CR proton and helium fluxes at high rigidities \citep{Evoli:2011id}, and might 
originate from a change in the turbulence power spectrum of the ISM. 
To investigate the implications of a change of this kind in $K(E)$,  we insert a break in the diffusion coefficient:
\begin{eqnarray}
\label{eq:broken_K}
K(E) =
\left\{
\begin{array}{rl}
& K_0  \,(E)^{-\delta_1} \,\,
\phantom{(E_{b}^K)^{+\delta_2-\delta_1}}
 E \leq E_{\rm b}^K, \\
& K_0 \, (E)^{-\delta_2}(E_{b}^K)^{-\Delta \delta} \,\,
E >E_{\rm b}^K,
\end{array}
\right.
\end{eqnarray}
The diffusion coefficient below the break energy is taken as in the standard case, i.e., $K_0$ and $\delta_1$ are those that refer either to the MED or the MAX case. The break acts above $E_b ^K$, where the spectral index changes by an amount $\Delta \delta= \delta_1 - \delta_2$. 
To investigate whether a spectral break in the diffusion coefficient could produce an effect similar to the one induced by
break in the injection spectrum, we compute the electron flux for a smooth SNR distribution 
by varying $\Delta \delta$ in the interval $0.1 - 0.6$ and compare it to the case where the injection spectrum of SNRs is a broken power law in Equation \eqref{eq:broken_Q}, with the standard $K(E)$ of Equation \eqref{eq:K}. 
We place the break $E_b^K$ at 60 GeV, in order to have a shape electron spectrum similar to the case for which we use a break in the injection spectrum. The result is shown in Figure \ref{fig:break_K}. 
%
We note that similar shifts in the diffusion coefficient or in the injection spectrum power laws 
$\Delta \delta \sim \Delta \gamma_{SNR}$ give electron fluxes described by different broken power laws. 
This is different from what one would expect for protons for instance, for which the flux is approximately $\Phi_p \propto Q(E)/K(E)$. 
In particular, the break required to fit the {\it Fermi}-LAT $e^+ + e^-$ spectrum ($\Delta \gamma_{SNR}$ = 0.4, see Table~\ref{tab:SNRsmoothbreak}), 
can not be due to a break in the diffusion coefficient, even with a very unlikely value for $\Delta \delta$. For example, the $\Delta \delta=0.6$ case would imply a very unlikely diffusion index $\delta_2=0.1$ above the break energy. Nevertheless,  the case with $\Delta \delta=0.6$ does not modify the electron flux sufficiently to obtain the hardening of the spectrum due to $\Delta \gamma_{SNR}$ = 0.4. 

We therefore conclude that a break at $E_b^Q=100$~GeV in the injection spectrum of a smooth Galactic SNR population can reproduce the potential break suggested by the {\it Fermi}-LAT $e^+ + e^-$ spectrum. On the other hand, a break in the diffusion coefficient is unable to reproduce the spectrum, even if the diffusion coefficient above the break is as hard as $K(E)\propto E^{-\delta_2} \sim E^{\;-0.1}$. We remark, however, that this result holds within the assumptions of our model, in which the diffusion coefficient is spatially independent. We cannot exclude {\it a priori} that a spatially inhomogeneous and/or anisotropic diffusion coefficient could induce a break in the observed spectrum for a single-power-law injection spectrum.



\section {Conclusions}  
\label{sec:conclusions}
The {\it Fermi}-LAT Collaboration has recently reported a new measurement of the inclusive $e^+ + e^-$ spectrum in the energy range between 7~GeV and 2~TeV, obtained with almost seven years of Pass 8 data \citep{Abdollahi:2017nat}. In this paper, we have explored several theoretical interpretations of this spectral measurement in terms of known sources: electrons and positrons emitted by primary sources, such as supernova remnants  and pulsar wind nebulae, or produced as a secondary CR component, due to the collision of protons and helium nuclei with the ISM.
The propagation of the leptons in the Galaxy has been modeled, including their large energy losses, by adopting the semi-analytical
model discussed in detail in \cite{2010A&A...524A..51D} and \cite{DiMauro:2014iia}.

We summarize our main results:
\begin{itemize}
\item The {\it Fermi}-LAT $e^+ + e^-$ spectrum is compatible with the sum of leptonic components arising from electrons produced by a smooth SNR population (distributed as in G15),  electrons and positrons coming from the PWNe in the ATNF catalog L04, and a secondary component. However, the PWNe emission turns out to exceed slightly the AMS-02 absolute positron flux.
\item When a prior on the positrons measured by AMS-02 is adopted, the higher-energy portion of the $e^+ + e^-$ spectrum does not reproduce the {\it Fermi}-LAT spectrum. This is the part of the spectrum where local sources (both for electrons and positrons) have
the largest impact.
\item When SNRs are separated into a far component (smoothly distributed as in G15) and a near component (SNR distance less than 0.7 kpc), where the near-component is populated by the SNRs present in Green's catalog, the agreement with the {\it Fermi}-LAT spectrum is significantly improved, including the high-energy tail. The improvement is especially visible in the case of a large confinement volume for CRs (the MAX model). However, once the electron emission from the brightest local SNRs, the Vela and the Cygnus Loop SNRs, is constrained from radio observations, the quality of the fit worsens. 
\item All these results have been obtained without invoking breaks in the spectral features of sources. A smooth distribution of SNRs with a break in the injection spectrum at $E^Q_b = 100 \pm 15$ GeV is the case that best reproduces the {\it Fermi}-LAT spectrum.
\item A spectral break in the diffusion coefficient is unable to reproduce the measured $e^+ + e^-$ spectrum.
\end{itemize}


In conclusion, the {\it Fermi}-LAT $e^+ + e^-$ spectrum can be reproduced either by local SNRs, as those present in Green's catalog and closer than about 1 kpc, or by a smooth distribution of sources endowed with a spectral break in the injection spectrum at about 100 GeV (at injection). In general, we find that the MAX propagation model performs better in reproducing the {\it Fermi}-LAT spectral measurement under all circumstances.

{\color{green} 
\medskip
}

\begin{acknowledgments}
The \textit{Fermi} LAT Collaboration acknowledges generous ongoing support
from a number of agencies and institutes that have supported both the
development and the operation of the LAT as well as scientific data analysis.
These include the National Aeronautics and Space Administration and the
Department of Energy in the United States, the Commissariat \`a l'Energie Atomique
and the Centre National de la Recherche Scientifique / Institut National de Physique
Nucl\'eaire et de Physique des Particules in France, the Agenzia Spaziale Italiana
and the Istituto Nazionale di Fisica Nucleare in Italy, the Ministry of Education,
Culture, Sports, Science and Technology (MEXT), High Energy Accelerator Research
Organization (KEK) and Japan Aerospace Exploration Agency (JAXA) in Japan, and
the K.~A.~Wallenberg Foundation, the Swedish Research Council and the
Swedish National Space Board in Sweden.
This work performed in part under DOE Contract DE-AC02-76SF00515.
Additional support for science analysis during the operations phase is gratefully acknowledged from the Istituto Nazionale di Astrofisica in Italy and the Centre National d'\'Etudes Spatiales in France.

This work is supported by the research grant {\sl Theoretical Astroparticle Physics} number 2012CPPYP7, funded under the program PRIN 2012 of the Ministero dell'Istruzione, Universit\`a e della Ricerca (MIUR) and by the research grants {\sl TAsP (Theoretical Astroparticle Physics)} and {\sl Fermi} funded by the Istituto Nazionale di Fisica Nucleare (INFN). This research was partially supported by a Grant from the GIF, the German-Israeli Foundation for Scientific Research and Development.
\end{acknowledgments}


\appendix
\section{Impact of adopting a SNR distribution with a spiral structure}\label{ap:sp}
\begin{figure*}[t]
\centering
\includegraphics[width=0.45\columnwidth]{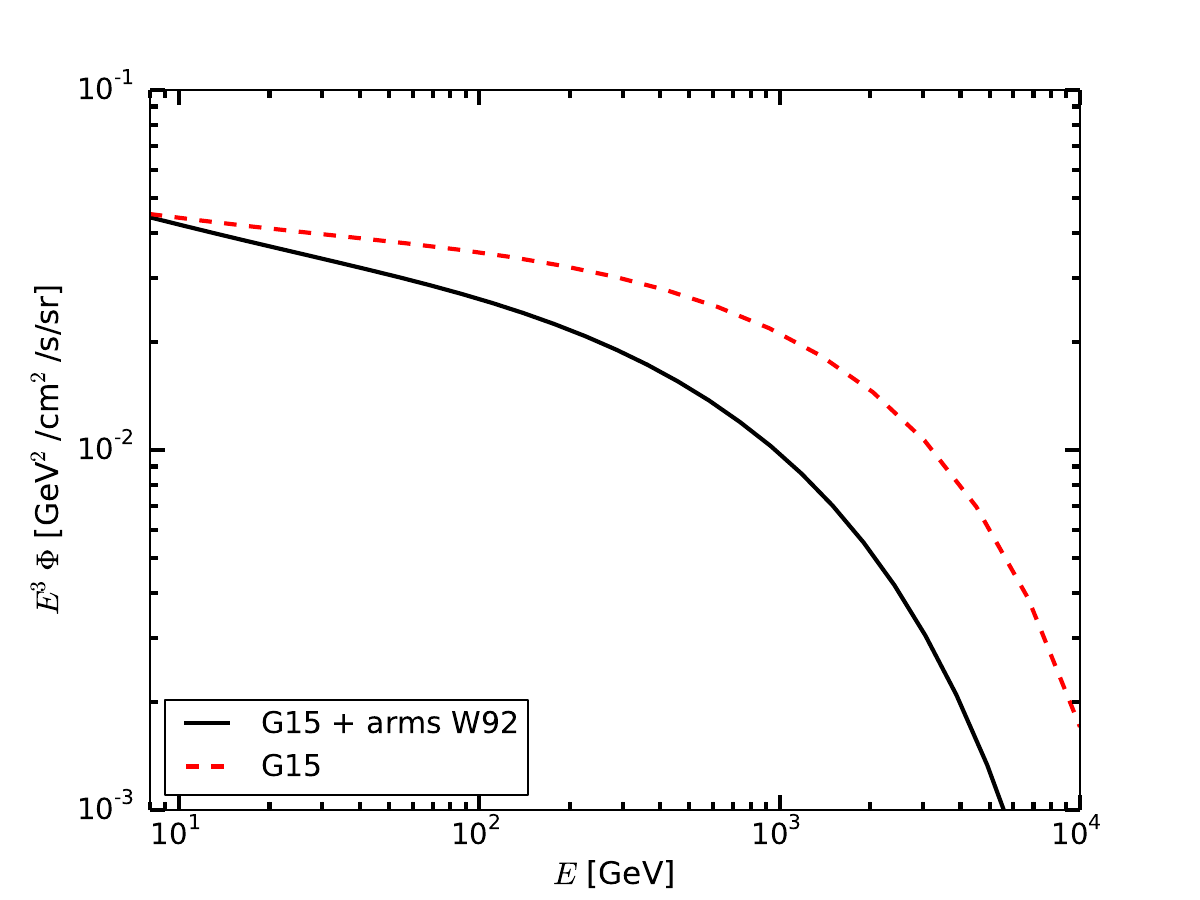}
\includegraphics[width=0.45\columnwidth]{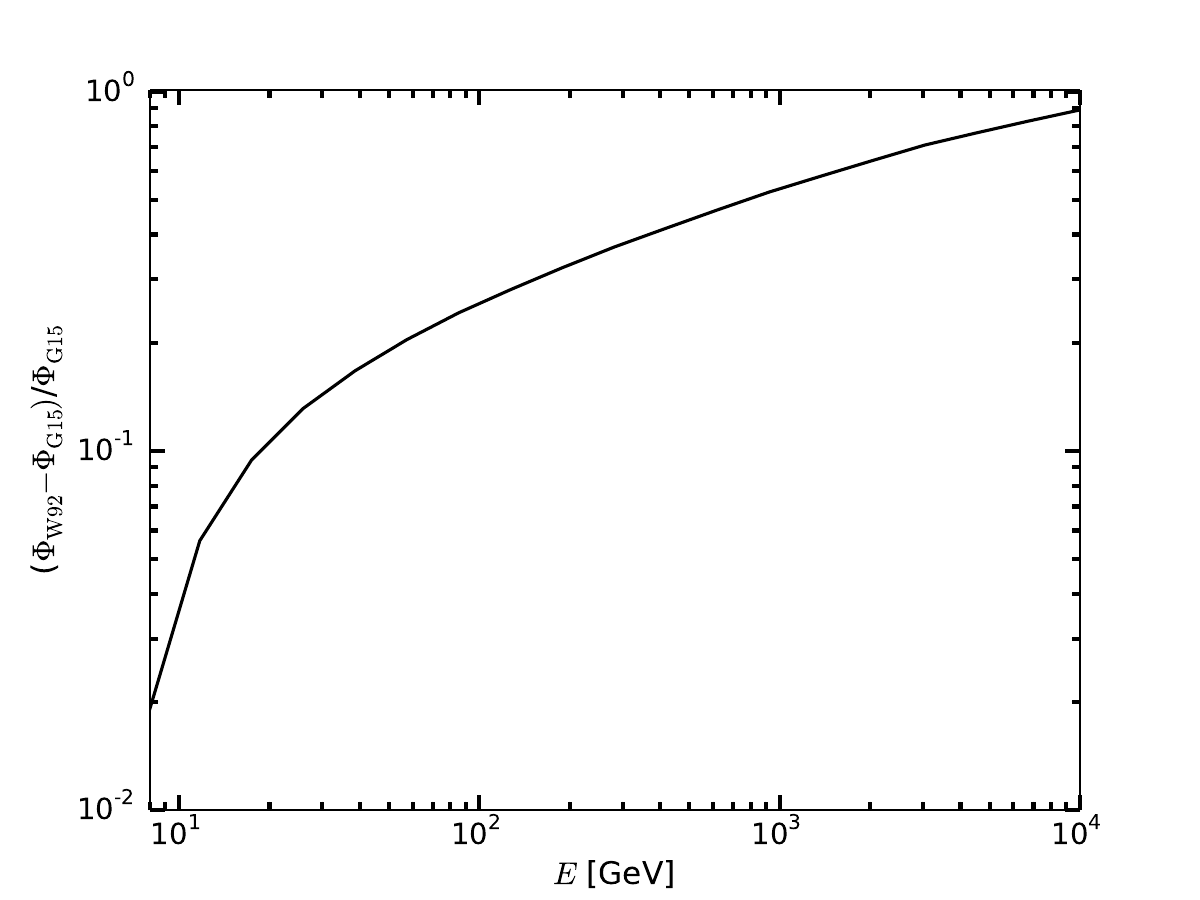}
\caption{
Effect of including a spiral pattern in the SNRs source distribution on the propagated electron flux. 
The fluxes are computed for the MAX propagation setup and using $\gamma_{\rm SNR}=2.4$. 
Left panel: electron fluxes for the G15 SNR distribution (red dashed line) and for the same distribution with the spiral pattern implemented according to the \cite{Wainscoat:1992pk} model, as described in the text (black solid line).
Right panel: relative difference $(\Phi_{\rm W92}- \Phi_{\rm G15})$ / $ \Phi_{\rm G15}$ as a function of energy. }
\label{fig:spiral}
\end{figure*}
The distributions of SNRs that are adopted everywhere in this paper (L04 and G15, see Sec.\ref{sec:distributions}) are azimuthally symmetric and therefore do not contain the spiral structure of the Milky Way. The impact of the presence of spiral arms on the cosmic rays fluxes at the Earth has been recently discussed by many authors \citep[see, e.g.,][]{2012A&A...547A.120E,Gaggero:2013rya,2014ApJ...782...34B,Werner:2014sya,Johannesson:2015qqi,2017JPhCS.837a2003K} by means of a fully three-dimensional description of the SNRs distribution function. In order to include the spiral arms presence in our semi-analytical technique, we model the distribution function as follows:
\begin{equation}
\rho(r, \phi,z)= \rho_0 \,f(r) \, e^{-\frac{|z|}{z_0}} \cdot S(r, \phi)
\end{equation} 
where $\rho_0$, $f(r)$ and $e^{-\frac{|z|}{z_0}}$ are discussed in Section \ref{sec:distributions}, while the function $S(r, \phi)$ describes the spiral pattern. Specifically, we adopt the four-arm structure described in \cite{Wainscoat:1992pk}, with the parameters provided by \cite{FaucherGiguere:2005ny}.
We compute the electron flux at Earth by solving  a more general version of Eq.~\eqref{eq:fluxsol} where any assumption on cylindrical symmetry is relaxed: 
\begin{eqnarray}\label{eq:fluxsol_spirals}
\Phi(\mathbf{x}_\odot, E&) &= \frac{v}{4\pi} \frac{\rho_0  \Gamma_{*}}{b(E)} \int dE_s \, Q(E_s) \nonumber
\\
&& \int dr_s d\phi_s \,\mathcal{G}(r_\odot, \phi_\odot, E \leftarrow r_s, \phi_s, E_s ) \,f(r_s)S(r_s, \phi_s) \nonumber
\\
&& \int dz_s\, \mathcal{G}_z(z_\odot, E \leftarrow z_s, E_s) \, e^{-\frac{|z_s|}{z_0}} 
 \end{eqnarray}
\
where the expressions for the Green functions $\mathcal{G}(r_\odot, \phi_\odot, E \leftarrow r_s, \phi_s, E_s )$ and $\mathcal{G}_z(z_\odot, E \leftarrow z_s, E_s)$ are the same as for the case without spirals, i.e. the ones derived in \cite{2010A&A...524A..51D}.
\\
\\
Our results are presented in Fig.~\ref{fig:spiral} for the MAX propagation models (we have checked that no significant difference in the spectral features appears if other propagation models are considered). In the left panel the flux for the G15 distribution with and without the spiral pattern suggested by the model \cite{Wainscoat:1992pk} are shown for the same spectral index $\gamma_{\rm SNR}=2.4$. The two curves are scaled to have the same normalization at $10$ GeV. We find that including the Galaxy spiral arms in our model produces a softening in the electron spectrum, in agreement with \cite{Gaggero:2013rya,2013JCAP...03..036D}. 
The softening is more pronounced at higher energies, as illustrated by the relative difference between the case with and without the spiral pattern in Fig.~\ref{fig:spiral} (right panel). This is due to the fact that the Earth sits in an inter-arm region and high-energy electrons lose more energy before they can reach it from the nearest arm, where the SNRs source distribution is peaked. 
\\
From these results we conclude that, given the importance of the softening at higher energies, the presence of the break in the $e^++e^-$ flux would be even more significant including the spiral pattern of the SNR source distribution.
\\
As a final remark, we emphasize that here we have implemented the spiral pattern only for the source distribution  and not for the energy losses or for the spatial diffusion coefficient. Modelling effects of this kind, albeit necessary to fully ascertain the impact of the presence of the spiral arms on our results, would require a fully numerical treatment of electron propagation and is therefore beyond the scope of the simple analytical model discussed in this work. 
\\




\bibliography{interpretation}

\end{document}